\documentclass[aps,pre,preprint,double-spaced,floatfix,showpacs]{revtex4}

\usepackage[dvips]{graphicx}
\usepackage{subfigure}
\usepackage{appendix}
\usepackage{epsfig}
\usepackage[english]{babel}
\usepackage{amsmath}
\usepackage{amssymb}
\usepackage{times}
\usepackage{color}

\begin{document}

\author{Jin W. Wang}

\affiliation{The James Franck Institute and The Department of Physics,
The University of Chicago, 929 East 57th Street, Chicago, Illinois 60637}

\title{Rim curvature anomaly in thin conical sheets revisited}

\date{\today}

\begin{abstract}
This paper revisits one of the puzzling behaviors in a developable cone (d-cone), the shape obtained by pushing a thin sheet into a circular container of radius $ R $ by a distance $ \eta $ \cite{cerda2}. The mean curvature was reported to vanish at the rim where the d-cone is supported \cite{liang2}. We investigate the ratio of the two principal curvatures versus sheet thickness $h$ over a wider dynamic range than was used previously, holding $ R $ and $ \eta $ fixed. Instead of tending towards 1 as suggested by previous work, the ratio scales as $(h/R)^{1/3}$. Thus the mean curvature does not vanish for very thin sheets as previously claimed. Moreover, we find that the normalized rim profile of radial curvature in a d-cone is identical to that in a ``c-cone" which is made by pushing a regular cone into a circular container. In both c-cones and d-cones, the ratio of the principal curvatures at the rim scales as $ (R/h)^{5/2}F/(YR^{2}) $, where $ F $ is the pushing force and $ Y $ is the Young's modulus. Scaling arguments and analytical solutions confirm the numerical results.

\end{abstract}


\maketitle
\section{INTRODUCTION}
\begin{figure}
\centering
\begin{tabular}{lr}
\epsfig{file=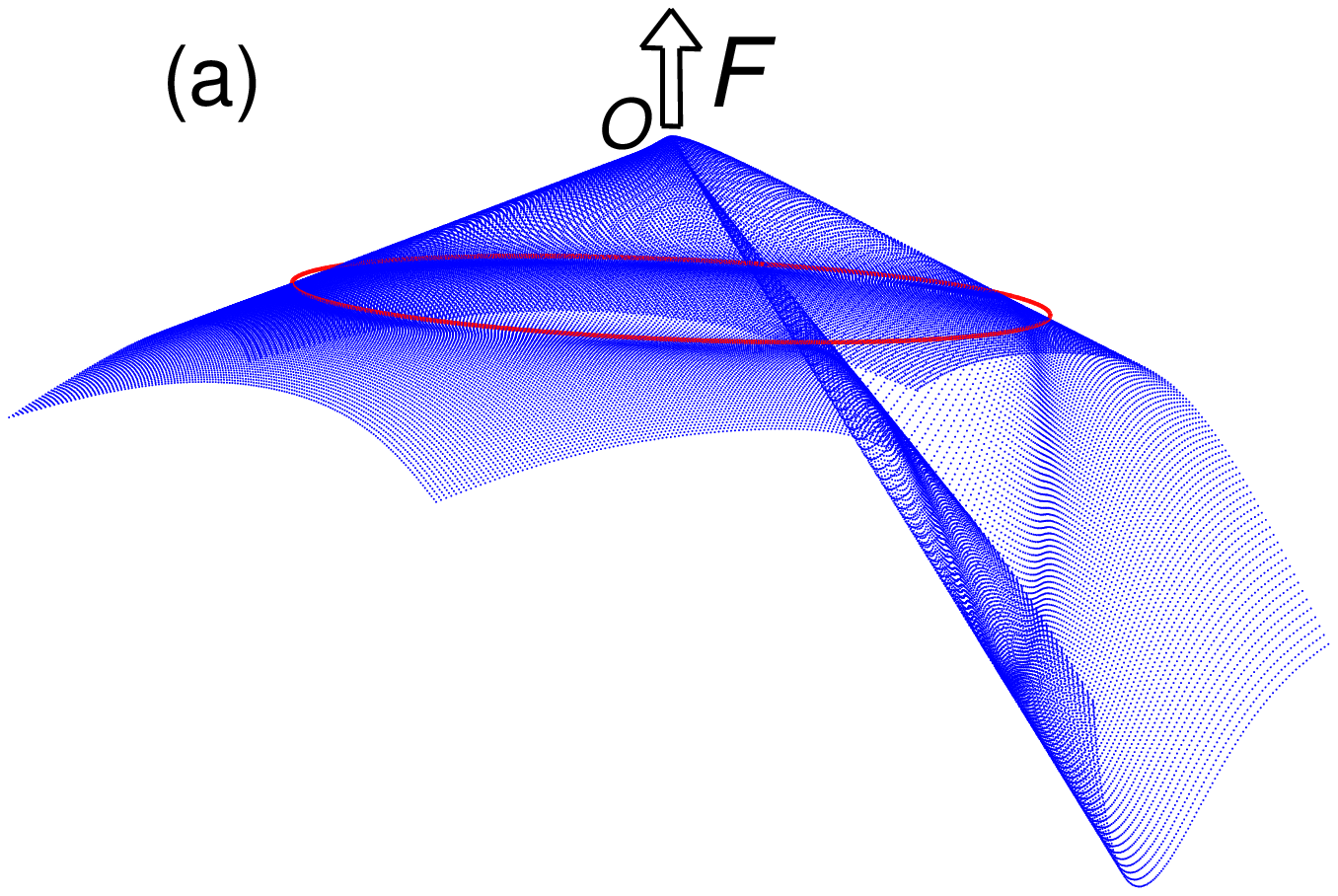,width=0.50\linewidth ,clip=} &
\epsfig{file=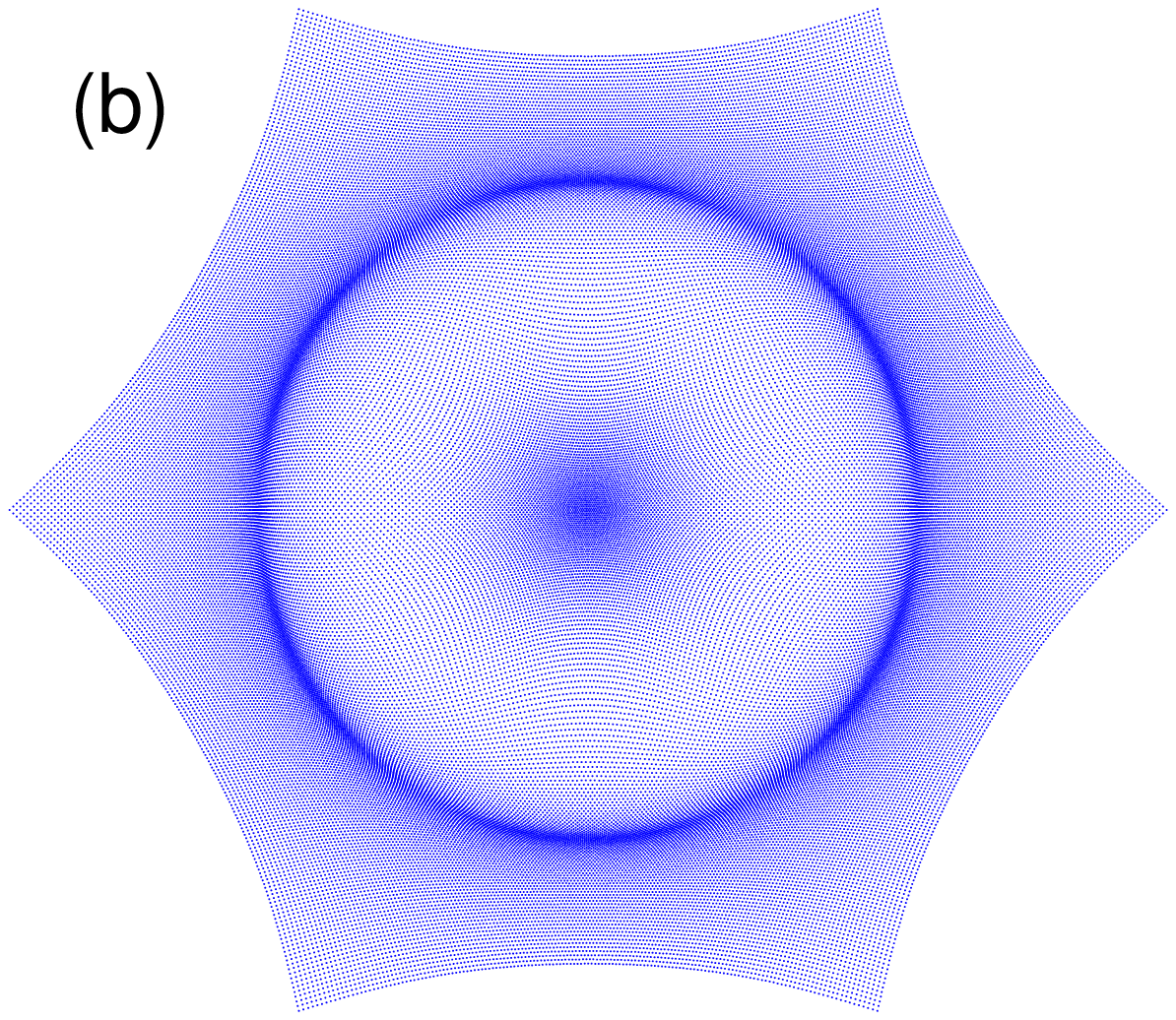,width=0.45\linewidth ,clip=} 
\end{tabular}
\caption{(Color online) (a) A typical simulated d-cone formed by pushing the center $ O $ of a hexagonal elastic sheet (Equations \ref{stretch_density} and \ref{bend_density}) against a circular container with concentrated force $ F $. The red solid line shows the rim of the container. The side length of the sheet $ \ell $ is $ 1.77 $ times the container radius $ R $. The thickness $ h=R/866 $, and the displacement of point $ O $ is a tenth of $ R $. For clarity, the vertical scale is expanded by about six times. (b) A variable lattice for simulating d-cones. It shows the material coordinates of the lattice points used in the simulation. The local lattice point densities at the rim and in the center are about $ 4.4 $ and $ 3.2 $ times the overall average point density, respectively. The average distance between adjacent lattice points is about $ R/88 $.}
\label{dcone}
\end{figure}

When a piece of thin sheet like paper crumples, it develops a network of two types of sharp structures: straight stretching ridges and pointlike vertices. Since thin sheets and membranes are very common in both natural and man-made structures at almost all length scales, crumpling has applications in a broad range of systems, such as graphene sheets \cite{fasolino2007}, carbon nanotubes \cite{Cao25112005, arias2008}, virus capsids \cite{lidmar1}, pollen grains \cite{Katifori27042010}, polymerized membranes \cite{Pocivavsek16052008}, and leaves \cite{Liang29122009}. The pointlike vertex singularity has mostly been studied via the simple realization known as the developable cone or d-cone, shown in Figure \ref{dcone}(a). A d-cone is the shape created by pushing the center of a thin sheet into a circular container of radius $ R $ by a distance $ \eta $ \cite{cerda2}. d-cones, stretching ridges, and crumpling in general have been studied extensively \cite{witten2, lobkovsky3, lobkovsky2, amar1, chaieb1,cerda3, chaieb2, cerda2, boudaoud1, mora1, didonna1, hamm1, Cerda17022004, Farmer2005509,blair2005,liang1, cerda1, andresen1, witten1, aharoni2010, Mellado2011}.  In certain cases it is possible to set bounds on the energy of singular structures \cite{Venkataramani2004, conti2008}, but to our knowledge such bounds have not been established for d-cones. Furthermore, while the scaling properties of stretching ridges have been determined analytically and numerically \cite{lobkovsky3, lobkovsky2, didonna1}, several phenomena in d-cones are still beyond our understanding \cite{witten1}.  One mysterious behavior of d-cones is the reported vanishing of mean curvature at the rim region where a d-cone is supported \cite{liang2}.

The deformation of a d-cone can be characterized by the deflection $ \epsilon\equiv \eta/R$ and its Young's modulus is denoted by $ Y $. $ r $ and $ \theta $ are defined as radial and angular components in the material coordinate system. $ C_{rr} $ and $ C_{\theta\theta} $ are the radial and azimuthal curvature respectively. In principle the shape of a d-cone is governed by the F\"{o}ppl-von K\'{a}rm\'{a}n equations\cite{amar1,landau1}, whose analytical solution is known only in a few special cases \cite{lobkovsky1,mansfield1}. Unfortunately, a d-cone is not one of these cases. However, it is energetically much cheaper for thin sheets to bend than to stretch, so a d-cone is asymptotically unstretched, and thence developable, except in the core region where it is pushed. Assuming that a d-cone is unstretched almost everywhere, Cerda and Mahadevan \cite{cerda3,cerda1} described the deformation in terms of the classical Elastica of Euler \cite{fraser1991} and obtained the shape of the d-cone by minimizing the bending energy. Numerical study \cite{liang1} has confirmed their finding.

Under the assumption that a d-cone is unstretched almost everywhere, the shape of the d-cone has zero radial curvature $ C_{rr} $ except in the core region, but the real shape must have nonzero $ C_{rr} $ at the rim to balance the normal force from the edge of the container\cite{liang2, wang1}. Liang and Witten \cite{liang2} reported a striking feature that within numerical accuracy, as the thickness of the sheet went to zero, the radial curvature $ C_{rr} $ and the azimuthal curvature $ C_{\theta\theta} $ were nearly equal and opposite at the rim, so that the mean curvature, defined as $ (C_{rr}+C_{\theta\theta})/2 $, nearly vanished there. They found that the feature was independent of container radius $ R $, thicknesses of the sheet $ h $, and deflection $ \epsilon $, but the circular symmetry of the container was indispensable. Since nonzero $ C_{rr} $ entails stretching, it should be necessary to consider the stretching in the rim region in order to understand this vanishing of mean curvature feature \cite{liang2}.

In the limit of thin sheets, the cost of stretching becomes prohibitive and the rim region with nonzero stretching should shrink to zero. In this limit, it seems plausible to treat the rim region as a boundary layer sandwiched by regions where the Elastica approach can still be applied. This type of boundary layer phenomenon appears in a wide variety of systems, such as the Pogorelov ring ridge \cite{landau1,pogorelov1988}, the ''minimal ridge" \cite{lobkovsky2}, and more recent work \cite{audoly1999,jin2000,lidmar1}.

Judging from its characteristics, this vanishing mean curvature phenomenon is nonlocal and purely geometric; thus, researchers have investigated the connection between this phenomenon and another nonlocal geometric constraint on surfaces, i.e. the Gauss-Bonnet theorem\cite{wang1,struik1}. This theorem requires that the sum of the integral of Gaussian curvature within a surface and the integral of geodesic curvature along its boundary remain a constant. For a d-cone, the integral of geodesic curvature along the boundary can be assumed a constant, so net Gaussian curvature in one region needs to be balanced by another region(s) of opposite Gaussian curvature. Since zero mean curvature at the rim means negative Gaussian curvature, researchers have suspected the negative Gaussian curvature at the rim is necessary to balance a net positive Gaussian curvatures in the core region\cite{wang1}. However, the integral of (negative) Gaussian curvature near the rim is almost completely compensated by that of two adjacent bands. Thus, the Gauss-Bonnet theorem offers no obvious explanation of the nonlocality implicit in the vanishing mean curvature phenomenon.

\begin{figure}
\centering
\begin{tabular}{lr}
\epsfig{file=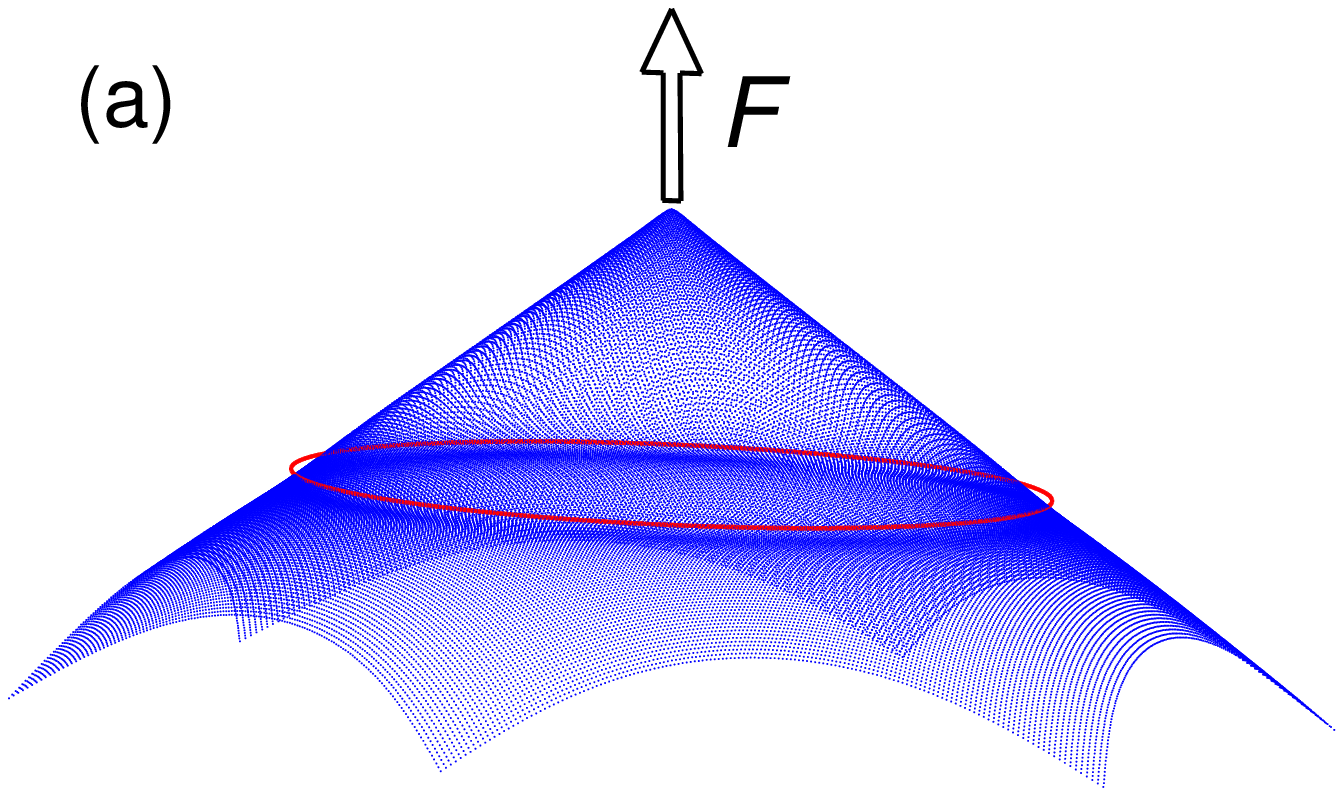,width=0.45\linewidth ,clip=} &
\epsfig{file=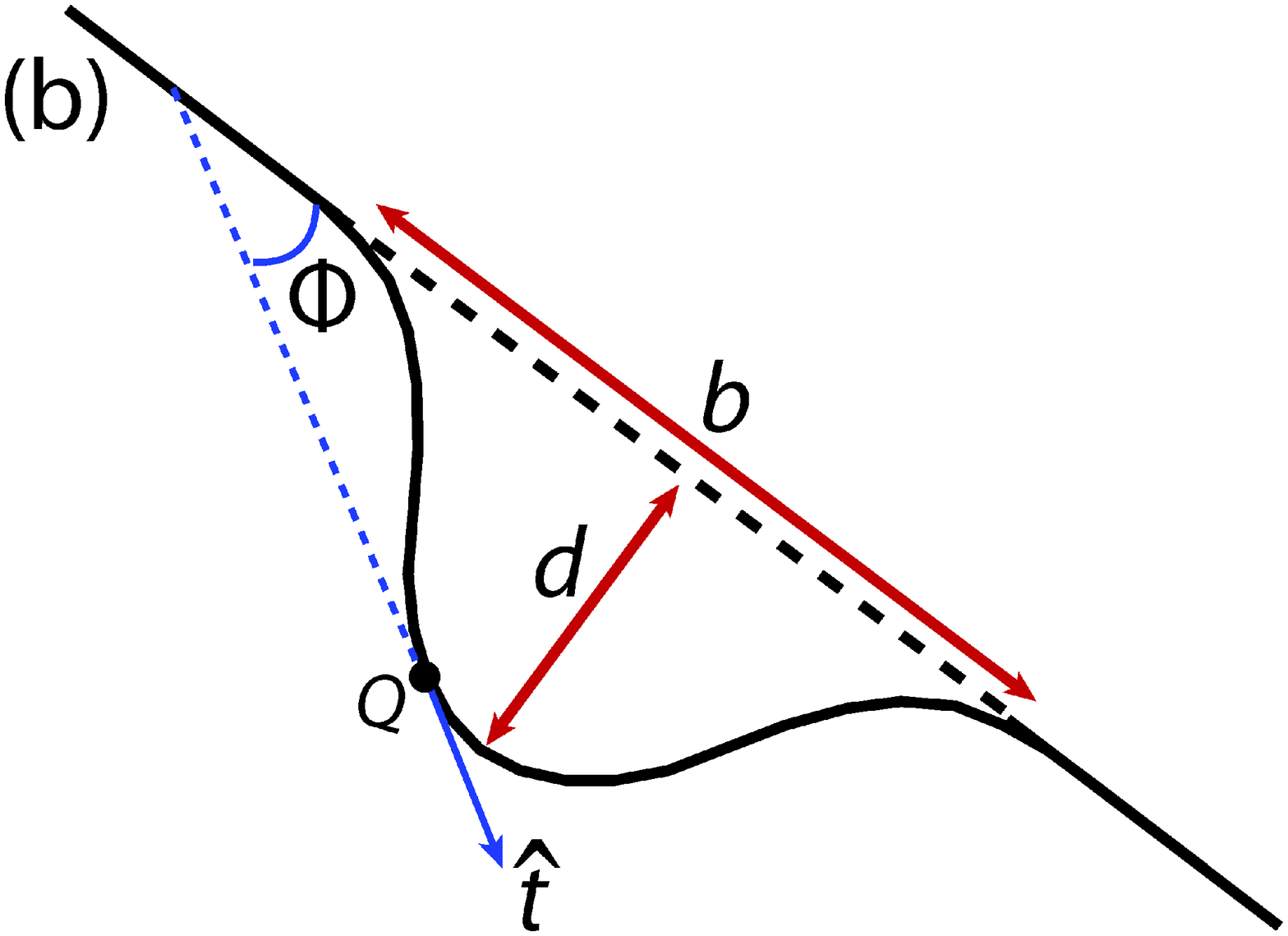,width=0.35\linewidth ,clip=} 
\end{tabular}
\caption{(Color online) (a) A typical simulated c-cone formed by pushing the center of a regular cone into a circular container with concentrated force $ F $. The opening angle of the cone is $ 168.58^\circ $, which translates to a deflection of $ 0.10 $. The thickness of the elastic sheet (Equations \ref{stretch_density} and \ref{bend_density}) $ h=R/866 $, and the height of the cone is about $ R/6 $. For clarity, the vertical scale is expanded by about ten times. (b) A schematic diagram of the local deformation near the rim along a meridian of the cone. The depth of the deformation is $ d $ and width is $ b $. For any point $ Q $, its tangent vector along the meridian is denoted by $ \hat{t} $, and the deviation of $ \hat{t} $ from the original unperturbed meridian is defined as $ \phi $.}
\label{ccone_shape}
\end{figure}

This paper investigates the ratio of the two principal curvatures $ |C_{rr}/C_{\theta\theta}| $ at the rim versus sheet thickness $h$ over a wider dynamic range than was used previously in Ref. \cite{liang2}, holding the deflection $ \epsilon $ and the container radius $ R $ fixed. The numerical models are specified in Sec. II. In Sec. III, we describe our numerical findings in detail. Instead of tending towards 1 as required by the vanishing of mean curvature, the ratio $ |C_{rr}/C_{\theta\theta}| $ at the rim goes below 1 and scales as $ (h/R)^{1/3} $. To better understand this power law, we study d-cone's close cousin ``c-cone" which is made by pushing a regular cone into a circular container, as seen in Figure \ref{ccone_shape}(a) . C-cones are simpler structures than d-cones. In a c-cone, deflection $ \epsilon $, pushing force $ F $, and thickness $ h $ are independent degrees of freedom. In both c-cones and d-cones, we find that $ |C_{rr}/C_{\theta\theta}|\propto (R/h)^{5/2}F/(YR^{2}) $ for fixed deflection $ \epsilon $. Moreover, given the same $ h $, $ R $, and $ \epsilon $, the normalized rim profiles of radial curvature are identical in a c-cone and a d-cone. To put these numerical findings on firmer grounds, in Sec. IV we give scaling arguments for both c-cones and d-cones. General solutions for symmetrically loaded conical shells are available \cite {love1944,turner1965}, and we use the proper boundary conditions to get the analytical solutions for c-cones. Both the scaling arguments and the analytical solutions confirm the numerical results. In Sec. V, we discuss the implications of our findings.

\section{NUMERICAL METHODS}
In principle, one could use standard finite element softwares such as Abaqus \cite{abaqus2009} to simulate thin sheets, but we find them difficult to adapt to the ultra-thin asymptotic behavior of d-cones that we want to study. We use an extended Seung-Nelson model \cite{wang2, didonna1} to cope with the singularity at the center of a d-cone and to better study the interaction between the sheet and the supporting container. The extended model simulates an elastic sheet by a triangular lattice with variable lattice spacing, so it has more adaptability than the original Seung-Nelson model \cite{seung1} which dictates a uniform lattice. This allows the extended model to concentrate lattice points as needed in regions of strong gradients. Our deformed lattice is shown in Figure \ref{dcone}(b). 

\begin{figure}
\centering
\begin{tabular}{lr}
\epsfig{file=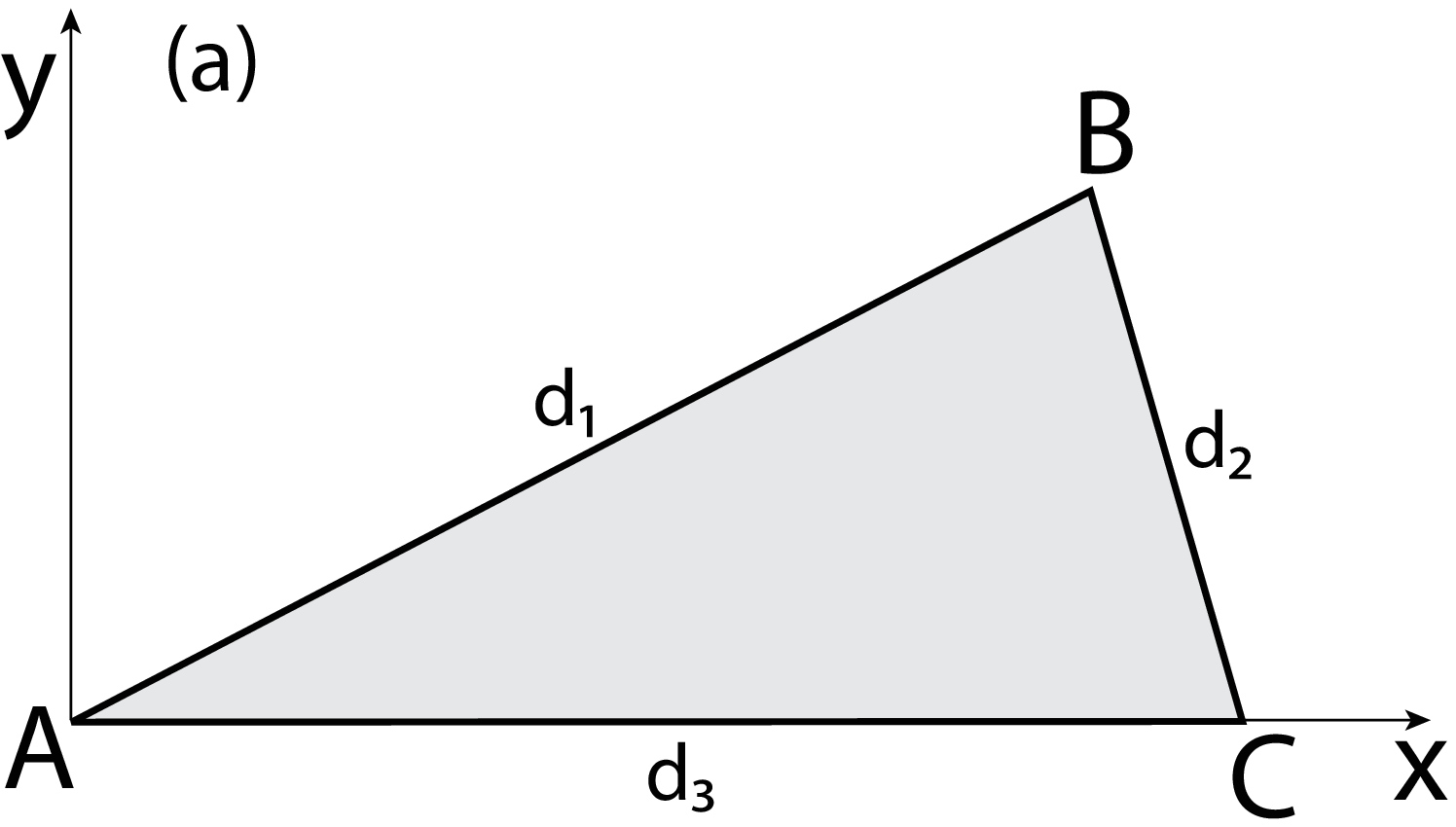,width=0.45\linewidth ,clip=} &
\epsfig{file=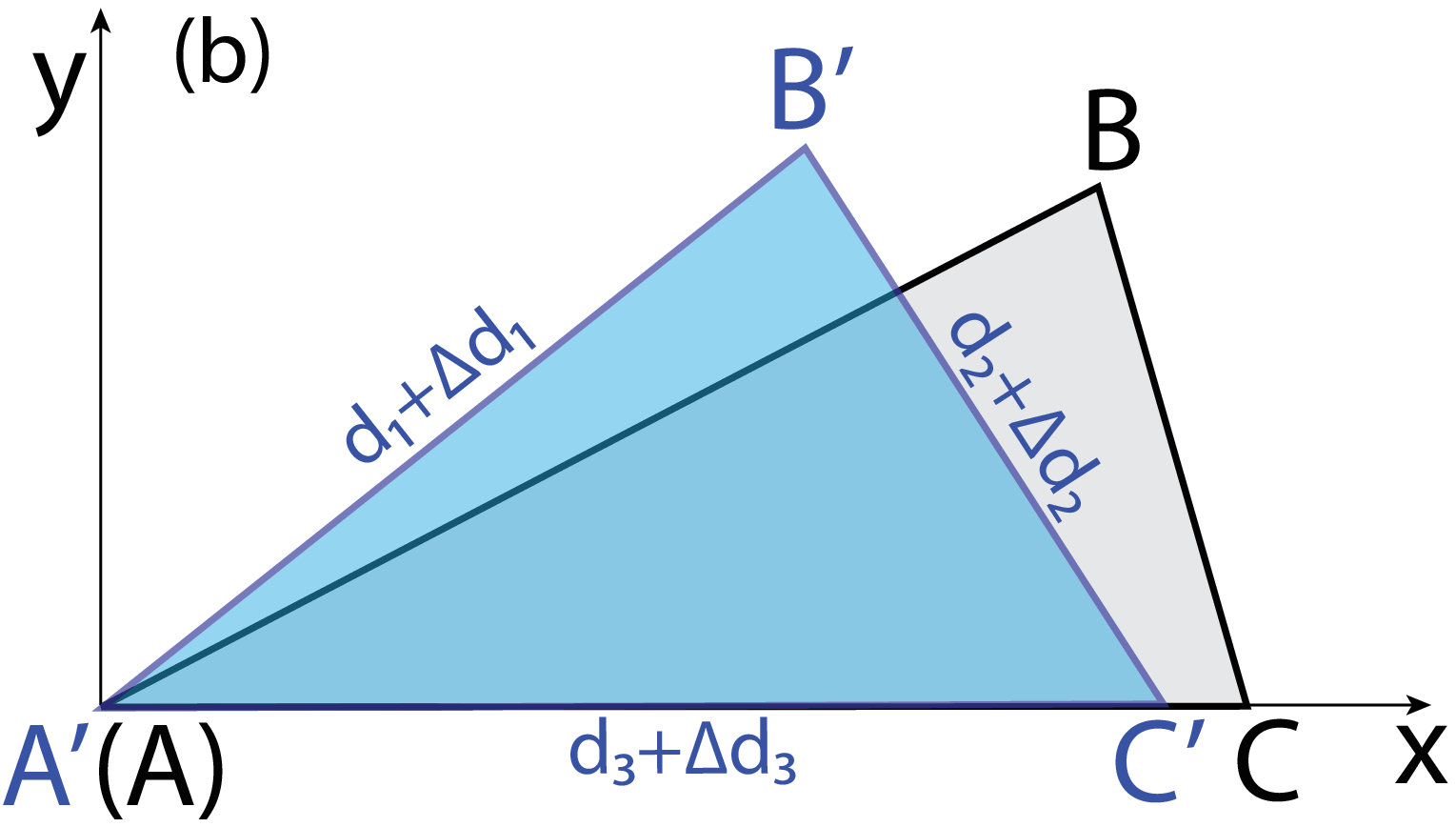,width=0.45\linewidth ,clip=} 
\end{tabular}
\caption{(Color online) (a) An arbitrary triangle $ ABC $ in its initial undeformed state. Its three edges have length $ d_{1} $, $ d_{2} $, and $ d_{3} $   (b) The triangle $ ABC $ is deformed into $ A'B'C' $. This stretching deformation can be captured either by the changes in the edge lengths: $ \Delta  d_1 $, $ \Delta  d_2 $, and $ \Delta  d_3 $, or by the strain tensor, which is assumed constant across the triangle.}
\label{strain}
\end{figure}

The total elastic energy of the sheet is the sum of stretching and bending energies on each triangle. On an arbitrary triangle the strain tensor and curvature tensor are assumed constant. The strain tensor $ \gamma $ is determined from the changes in the edge lengths, as seen in Figure \ref{strain}, and the curvature tensor $ C $ from the dihedral angles between the given triangle and its three adjoining triangles as shown in Figure \ref{bend_fig}. The specific transformation formulas are derived in Appendix A. Once we know the strain and curvature tensors, we obtain the corresponding energy densities $E_S$ and $E_B$ via the conventional equations of elasticity \cite{landau1,mansfield1}:
\begin{align}
&E_{S}=\frac{hY}{2(1-\nu^{2})}[(\text{Tr}(\gamma))^{2}+2(\nu-1)\text{Det}(\gamma)], \label{stretch_density} \\
&E_{B}=\frac{1}{2}\kappa(\text{Tr}(C))^{2}+\kappa_{G}\text{Det}(C),   \label{bend_density}
\end{align}
where $ \nu $ is the Poisson's ratio, $ \kappa =Yh^3/(12(1-\nu^2))$ is the bending rigidity, and $ \kappa_{G} $ is the Gaussian bending rigidity. The total elastic energy of the sheet is taken as the sum of the energy density for each triangle times its undeformed area.

\begin{figure}[h!]
  \begin{center}
  \includegraphics[width=0.5\textwidth]{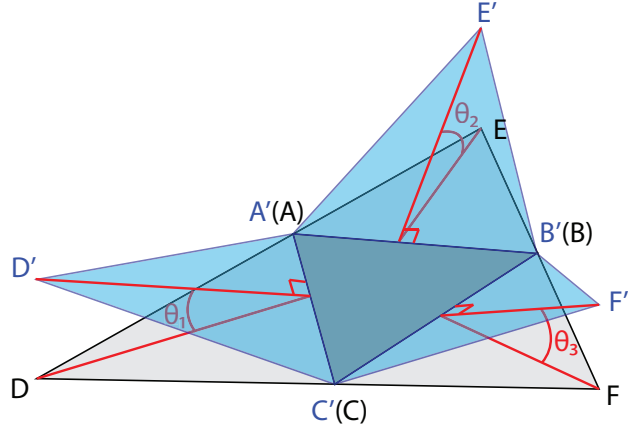}
  \caption{(Color online) An arbitrary triangle $ ABC $ and its three adjoining triangles in both the initial undeformed stated and the deformed state in a local coordinate system. This bending deformation can be captured either by the three dihedral angles $ \theta_1 $, $ \theta_2 $, and $ \theta_3 $ between $ ABC $ and its three neighbors or by the curvature tensor, which is assumed constant across the four triangles.}
  \label{bend_fig}
  \end{center}
\end{figure}

\begin{figure}[h!]
  \begin{center}
  \includegraphics[width=0.5\textwidth]{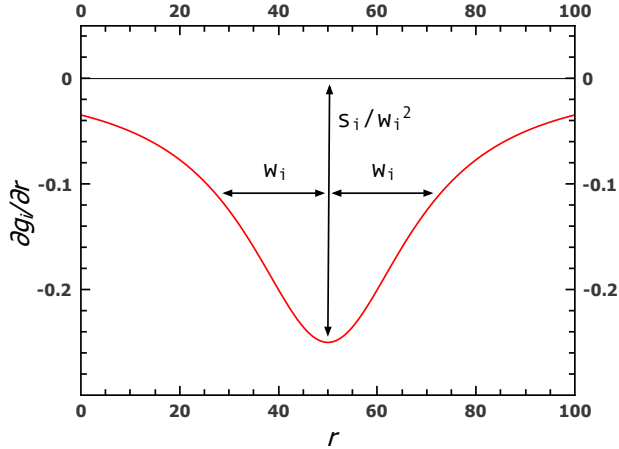}
  \caption{(Color online) $ \frac{\partial g_i}{\partial r} $ vs. $ r $ when $ r_i=50 $, $ w_i=20 $, and $ s_i=-100 $.  }
  \label{mapping_function}
  \end{center}
\end{figure}

A variable lattice for a d-cone is created in two stages \cite{wang2, didonna1}. First a uniform triangular lattice of spacing $ a $ is used to span the desired area. The lattice spacing $ a $ is the distance between adjacent lattice points. Then this lattice is mapped to the desired nonuniform lattice e.g. Figure \ref{dcone}(b). Let the point density of the initial uniform lattice be $ \rho_{init} $. For our purposes we may choose a mapping with radial symmetry. For a point $ (r,\theta) $ in the uniform lattice, we transform $ r $ so its new position in the variable lattice is $ (\tilde{r}(r),\theta) $.  Then the local point density in the variable lattice is $ \rho_{init}/(\frac{\partial \tilde{r}}{\partial r}\frac{\tilde{r}}{r}) $. In practice, we choose $ \tilde{r} $ to be a function form of $ \tilde{r}=r+\sum \limits_{i} g_{i}(r, r_i, w_i, s_i) $, where
\begin{gather*}
g_{i}(r, r_i, w_i, s_i)=\frac{s_i}{w_i}[\arctan(\frac{r-r_i}{w_i})-\arctan(-\frac{r_i}{w_i})] ,\\
\frac{\partial g_{i}(r, r_i, w_i, s_i)}{\partial r}=\frac{s_i}{(r-r_i)^2+w_i^2} ,
\end{gather*}
and the index $ i $ labels the regions where local point density needs adjustment. In Figure \ref{dcone}(b) there are four such regions: the center region, the rim region, the region between the center and the rim, and the outer region. In the center region and the rim region the point densities are increased while in the two other larger regions the point densities are reduced. The overall average point density remains close to that of the uniform lattice before the density adjustment. As shown in Figure \ref{mapping_function}, the graph of $ \frac{\partial g_{i}}{\partial r} $ is a simple U-shaped curve centered around $ r_i $, and we can control its width through $ w_i $ and depth through $ s_i $.

The resulting nonuniform lattice defines the initial undeformed state of Figure \ref{strain} for each triangle in the nonuniform lattice. Thus the lattice positions defined by this map constitute the state of zero stretching energy. By construction, these positions lie in a plane, so that each triangle also has zero curvature energy as defined by Figure \ref{bend_fig}.

Creating the nonuniform lattice for a c-cone requires two additional steps. To simulate a c-cone with opening angle equal to $ 2\theta_0 $, we make a cut along the radial line $ \theta=0 $, and map every lattice point through the transformation $ r'=\tilde{r} $ and $ \theta'=sin(\theta_0)\theta $. The last step is to join the two free boundary lines. That is, we identify each point on the free radial boundary line with its counterpart on the other free boundary line, so that the two lattice positions are constrained to occupy the same spatial position in the simulation.

The constraining container rim and the pushing force are simulated in almost the same way as previously used in Ref. \cite{liang1,wang1}. The rim is in the $ x-y $ plane and its shape is determined by the equation $ x^2+y^2=R^2 $, where $ R $ is the radius of the container. We introduce an external potential to implement the geometric rim constraint $ (\sqrt{x^2+y^2}-R)^2+z^2 \neq 0$ for all points in the sheet. To implement such a constraint for a discrete lattice we must assure that every lattice point remains a distance of order $ a $ from the rim line. For numerical tractability we assure this by adding an external potential felt by all lattice points that maintain the required separation while having negligible effect on more distant points. We find empirically that a short range $ r^{-8} $ potential is adequate. More specifically, we implement the repulsive normal force from the rim by introducing a potential of the form 
\begin{equation}
U_{\text{rim}}=\sum \limits_{j} C_{p}/[(\sqrt{x_j^2+y_j^2}-R)^2+z_j^2]^4 \, , \nonumber
\end{equation}where $ C_p $ is a constant, $ (x_j, y_j, z_j) $ is the coordinate of the $ j $th lattice point, and the summation is over the whole lattice. The value of $ C_p $ is chosen so that the shortest distance between the lattice points and the rim is approximately the local lattice spacing in the radial direction. The potential due to the central pushing force is 
\begin{equation}
U_{\text{force}}(x_1, y_1, z_1)=- (z_1+a)G(x_1,y_1)F \, . \nonumber
\end{equation}
Here $ (x_1, y_1, z_1) $ is the coordinate of the lattice point in the center, $ F $ is effectively the magnitude of the pushing force, and the function $ G(x_1, y_1) $ is given by
\begin{equation}
G(x_1, y_1)=[(1+(x_1/\xi)^2)(1+(y_1/\xi)^2)]^{-1} \, , \nonumber
\end{equation}
where $ \xi $ is a constant of order $ 0.1a $. This $ G(x_1, y_1) $ is introduced to make sure that when the sheet is being pushed the lattice point in the center does not stray away from the axis of the cylindrical container, i.e., $(x_1,y_1)\simeq (0,0)$. 

The total energy of the system is the sum of the total elastic energy in the sheet and the potential energies due to the rim and the applied pushing force. The conjugate gradient algorithm\cite{press1} is used to minimize the total energy as a function of the coordinates of all lattice points to get the final shape of the sheet \cite{janosi1998,liang1,wang1}. To verify that the energy of the final configuration is at a global minimum, we can move each lattice point away from its equilibrium position by a random amount in a random direction and see if the energy minimization process will bring the system back to its original state; the magnitudes of the random displacements introduced are usually much less than the local lattice spacing. We can also check how sensitive the final configuration is to the starting configuration. A high sensitivity generally signals the thickness of the sheet is too small for the current lattice to simulate and the results are likely unreliable.

This model faithfully represents continuum sheets provided that the radii of curvature are everywhere much larger than the local lattice spacing. This limitation restricts the values of deflection $ \epsilon $ to be less than or equal to $ 0.15 $ in our simulations. Ref. \cite{wang2} reports further details about this simulation technique. The simulation program using the nonuniform lattice has been validated to show uniform elastic behavior for states of planar stress and for cylinders.  It has been validated against uniform lattices for d-cones of thickness $h$ that can be simulated by both methods.  Ref. \cite{wang2} also provides further simulated sheets, as well as our relaxation protocol and timing information.

Our computer program is sequential and the bottleneck of the simulation is the CPU speed. A typical d-cone or c-cone is simulated with $ 67,951 $ lattice points, and it usually takes more than one month to finish the energy minimization process on a $ 2.67 $ $ GHz $  Intel Core $ i7 $ processor. Each processor usually runs only one instance of the program at a time, but we have up to $ 40 $ processors to run multiple instances with different parameters simultaneously.

Once the final configuration of a d-cone is obtained, the curvature tensor in each triangle can be determined using the procedure stated earlier. We then solve the characteristic equation of each curvature tensor to get the principal curvatures $ C_{rr} $ and $ C_{\theta\theta} $. To find the radial profile of $ C_{rr} $, we pick a circular sector with angle $ 0.05 $ and for each triangle within this sector, its $ C_{rr} $ is associated with the radial coordinate of its center. As shown later, $ C_{rr} $ reaches its maximum at the rim where the interaction between the sheet and the container is the strongest. Within each sector we average the four largest values of $ C_{rr} $ near the peak as the rim value of $ C_{rr} $. The four corresponding $ C_{\theta\theta} $ are averaged to get the rim value of $ C_{\theta\theta} $. The rim curvatures depend on the angular separation between the center line of the circular sector and that of the buckled region. The angular separation is $ 5\pi/6 $ for the rim profiles of $ C_{rr} $ and rim values of $ |C_{rr}/C_{\theta\theta}| $ reported in subsequent sections. The estimated percentage uncertainty of the reported rim values of $ |C_{rr}/C_{\theta\theta}| $ due to this angular dependence is about $ 10\% $ for the whole range of thickness considered. A better approach is to average the rim $ C_{rr} $ and $ C_{\theta\theta} $ across multiple sectors, but this will have no noticeable impact on our results. Our tests also show that different initial configurations will cause $ |C_{rr}/C_{\theta\theta}| $ at the rim to change by less than $ 4\% $ for the range of thickness considered here. If we assume these two sources of uncertainties are independent, the total percentage uncertainty for the reported rim values of $ |C_{rr}/C_{\theta\theta}| $ should be about $ 11\% $. The results for c-cones are determined in the same way and have the same level of uncertainties as in the d-cone data.

\section{NUMERICAL RESULTS}
\begin{figure}[h!]
  \begin{center}
  \includegraphics[width=0.5\textwidth]{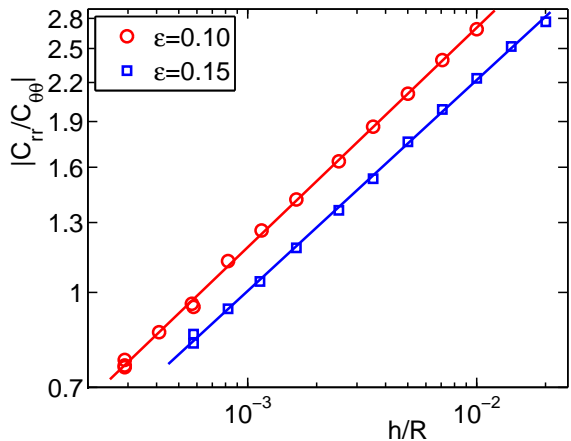}
  \caption{(Color online) The ratio of the two principal curvatures at the rim of a d-cone as a function of the relative thickness of the sheet for $ \epsilon=0.10 $ and $ \epsilon=0.15 $. It clearly shows that the ratio is less than one for very thin sheets and the ratio keeps decreasing as the sheet gets thinner. The slopes of the fitted lines are 0.36 and 0.34 for $ \epsilon=0.10 $ and $ \epsilon=0.15 $ respectively.}
  \label{ratio_dcone_fig}
  \end{center}
\end{figure}

Figure \ref{ratio_dcone_fig} shows $ |C_{rr}/C_{\theta\theta}| $ at the rim of d-cones versus relative thickness $h/R$ for two different values of $ \epsilon $. Unless explicitly stated otherwise, $ R $ and Young's modulus $ Y $ are assumed to be constant. For both $ \epsilon=0.10 $ and $ \epsilon=0.15 $, $ |C_{rr}/C_{\theta\theta}| $ goes below 1 as $h/R$ is sufficiently thin, and it does not show any sign of leveling off as it reaches as low as 0.76 in the thinnest sheets simulated. More strikingly, for each fixed $ \epsilon $, $ |C_{rr}/C_{\theta\theta}| $ scales as $ (h/R)^{1/3} $. This clearly contradicts the previous observation of vanishing mean curvature which requires $ |C_{rr}/C_{\theta\theta}| $ to stay at 1 when the sheet gets very thin. To resolve this contradiction, we need a better understanding of how $ |C_{rr}/C_{\theta\theta}| $ at the rim responds to changes in $ h $ and $ F $. However, in d-cones there is a one-to-one correspondence between $ h $ and $ F $ for a fixed $ \epsilon $. To gain more flexibility and to explore the generality of this feature, we study d-cone's close relative, the c-cone.

\begin{figure}[h!]
  \begin{center}
  \includegraphics[width=0.5\textwidth]{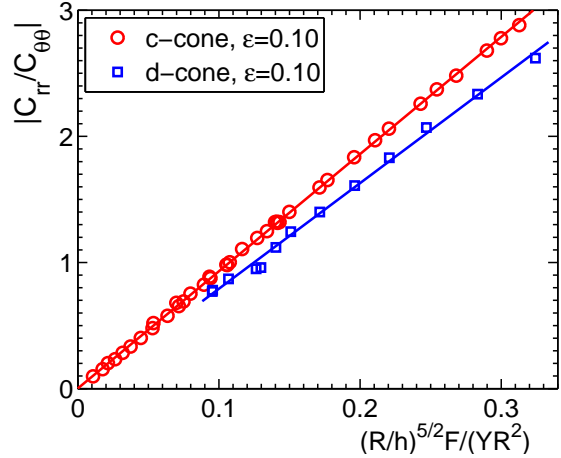}
  \caption{(Color online) $ |C_{rr}/C_{\theta\theta}|  $ at the rim as a function of $ (R/h)^{5/2}F/(YR^{2}) $ for both  c-cones and d-cones. $ h $ and $ F $ are changed independently in the c-cone data. For a specific $ h $, if we denote $ F_d $ as the center force needed in the d-cone to give the $ \epsilon $ of the c-cone, namely 0.1, the pushing forces used in the corresponding c-cone simulations may vary from $ 0.12F_d $ to $ 1.5F_d $.}
  \label{crr_over_ctt_ratio_ccone_dcone}
  \end{center}
\end{figure}

\begin{figure}[h!]
  \begin{center}
  \includegraphics[width=0.5\textwidth]{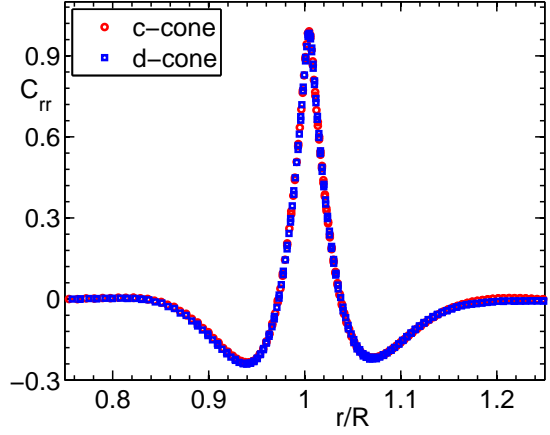}
  \caption{(Color online) Radial profiles of normalized $ C_{rr} $ in both a c-cone and a d-cone. The c-cone and the d-cone have the same $ \epsilon $, $ h $, $ R $, and $ F $. Each profile is normalized by dividing the vertical scale by its corresponding peak value. There is no normalization for the horizontal scale. As stated in Section II, for the d-cone the angular separation between the buckled region and the radial sector used to generate the radial profile is $ 5\pi/6 $. }
  \label{crr_vs_dcone_radial}
  \end{center}
\end{figure}

In c-cones, for a fixed $ \epsilon=0.10 $, while $ h $ and $ F $ are changed independently we find that $ |C_{rr}/C_{\theta\theta}| $ at the rim scales as $ (R/h)^{5/2}F/(YR^{2}) $, as shown in Figure \ref{crr_over_ctt_ratio_ccone_dcone}. This scaling law is justified in Section IV. In d-cones, even though $ h $ and $ F $ are interdependent, the same scaling law  also holds. Moreover, Figure \ref{crr_vs_dcone_radial} demonstrates that for the same set of $ h $, $ F $, and $ \epsilon $ the radial profile of normalized $ C_{rr} $ near the rim is the same in the d-cone as in the c-cone. These two findings together suggest very strongly that the behavior of $ C_{rr} $ in the rim region of a d-cone is identical to that in a c-cone, and should be explained by the same mechanism.

\begin{figure}[h!]
  \begin{center}
  \includegraphics[width=0.5\textwidth]{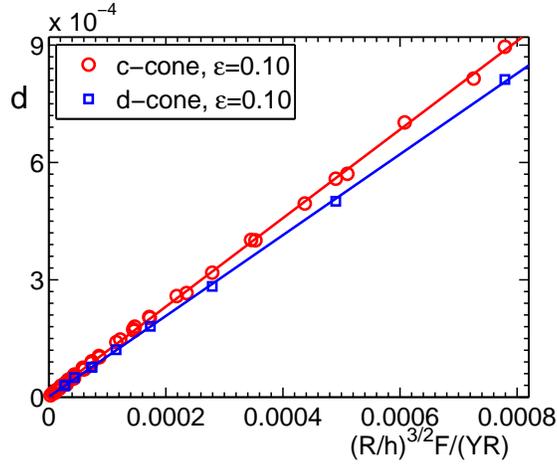}
  \caption{(Color online) The maximum deviation $ d $ of the local deformation near the rim as a function of $ (R/h)^{3/2}F/(YR) $. The constant container radius $ R $ is set to be the unit length. In the c-cone data $ h $ and $ F $ are changed independently as in Figure \ref{crr_over_ctt_ratio_ccone_dcone}. }
  \label{d_vs_F_h_cd_cone}
  \end{center}
\end{figure}

\begin{figure}[h!]
  \begin{center}
  \includegraphics[width=0.5\textwidth]{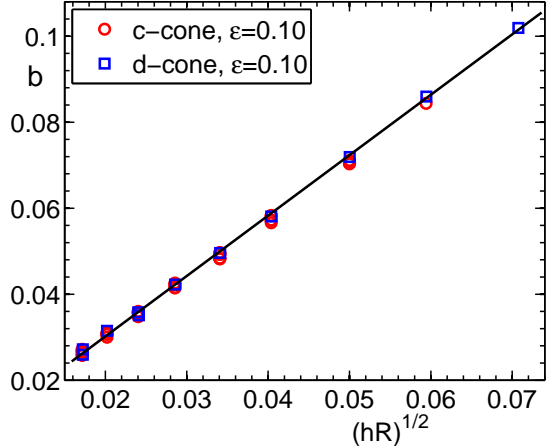}
  \caption{(Color online) The width $ b $ of the local deformation near the rim as a function of $ \sqrt{hR}$. More specifically for the data shown here, width $ b $ is the full-width-at-half-maximum of the radial profile of $ C_{rr} $ as shown in Figure \ref{crr_vs_dcone_radial}. The constant container radius $ R $ is set to be the unit length. Again, $ h $ and $ F $ are changed independently in the c-cone data  as in Figure \ref{crr_over_ctt_ratio_ccone_dcone}. This plot shows that $ b $ does not depend on $ F $ in c-cones.}
  \label{b_vs_h_R_cd_cone}
  \end{center}
\end{figure}

To shed some light on the underlying mechanism, we investigate the detailed deformation of the sheets near the supporting rim, as sketched in Figure \ref{ccone_shape}(b). Let us denote the maximum deviation from the straight radial line as  $ d $ and the width of the deformation as $ b $. Figure \ref{d_vs_F_h_cd_cone} shows that $ d $ has a linear response to $ F $ and scales as $ (R/h)^{3/2}F/(YR) $. From Figure \ref{b_vs_h_R_cd_cone}, we can see that $ b $ is independent of $ F $ and scales as $ \sqrt[]{hR} $. It's worth noting that for the same $ h $ and within numerical accuracy $ b $ is exactly the same in a d-cone as in a c-cone. It scales in the same way as the width of a Pogorelov ring ridge formed by pushing a convex thin shell with a large inward concentrated normal force \cite{landau1,pogorelov1988}. Our scaling arguments in the next section will closely follow how the scaling properties of the Pogorelov ring ridge are derived.

\section{SCALING ARGUMENTS AND ANALYTICAL SOLUTIONS}
\subsubsection{Scaling arguments}

Scaling arguments can be constructed to determine the dependence of ratio $ |C_{rr}/C_{\theta\theta}| $, the maximum deviation $ d $, and width $ b $ on $ h $, $ R $, and $ F $. Near the rim of a c-cone or a d-cone, the energy contribution of the local deformation includes both bending and stretching energy. Let us first find how each of them scales with $ d $ and $ b $. 

To evaluate the bending energy component, we start with the principal curvatures. We denote quantities unperturbed by the rim force by an overbar, e.g., $ \bar{C}_{\theta\theta} $. We denote changes induced by the the rim force by a $ \Delta $, e.g., $ \Delta C_{\theta\theta} $. Let $ \phi $ be the deviation of the tangent vector away from the original direction along a meridian as shown in Figure \ref{ccone_shape}(b). In the deformed region $ \phi\sim d/b$ and $ C_{rr}=\Delta C_{rr}\sim  \phi/b \sim d/b^{2}$ . We also have $ \bar{C}_{\theta\theta}\sim 1/R $, and radius of the azimuthal curvature $ \bar{R}_{\theta\theta}=1/\bar{C}_{\theta\theta} \sim R $. Due to the local deformation, $ \Delta  R_{\theta\theta} \sim -d $, so $ \Delta C_{\theta\theta} \sim d/{{\bar{R}_{\theta\theta}} }^{2} $, or $ d/R^2 $. Since $ b \ll R $ as argued below, $ \Delta C_{\theta\theta}$ is much less than $\Delta C_{rr} $ and can be safely ignored. Locally there is a contribution to the $ \kappa_G $ part of the bending energy of Equation \ref{bend_density}. The change in this energy induced by the rim force is of order $\kappa_G \Delta C_{rr} C_{\theta\theta} $. However, this $ \kappa_G $ energy must integrate to zero because of the Gauss-Bonnet theorem. Thus the bending energy $ U_{B, rim} $ due to the local deformation near the rim (over an area $ \sim bR $) is:
\begin{equation}
U_{B,rim} \sim \kappa {C_{rr}}^2 bR \sim \kappa Rd^2/b^3.\label{bend_energy_rim} 
\end{equation}

For the stretching component, $ \gamma_{\theta\theta} \sim d/R $ and $ \gamma_{rr} $ is negligible here \footnote{Due to the local deformation, the length of the meridian within a distance of $ b $ increases by $u_r \sim \sqrt[•]{d^2+b^2}-b \sim d^2/b $, but the stretching due to this elongation is spread over a distance much larger than $ b $. This can be shown through the following exercise. For the cone shown in Figure \ref{ccone_shape}(a), let us remove the top part that is above the rim, fix the outer edge, and pull the top edge of the remaining cone along the meridian away from the outer edge by a distance of $ u_r $. Then $ \gamma_{rr} \sim u_r/X $ and $ \gamma_{\theta\theta} \sim u_r/R $, where $ X $ is the decay length of $ \gamma_{rr} $. The resulting total stretching energy will be $ Eh[(u_r/X)^2+(u_r/R)^2]RX $. Minimizing this energy gives $ X\sim R $.}. Thus the stretching energy contribution near the rim is: 
\begin{equation}
U_{S,rim} \sim hY{\gamma_{\theta\theta}}^2 bR \sim hYbd^2/R.\label{stretch_energy_rim} 
\end{equation}

Minimizing $U_{B,rim}+U_{S,rim} $ gives $ b \sim \sqrt[•]{hR} $.  Plugging this back into Equations \ref{bend_energy_rim} and \ref{stretch_energy_rim}, the total elastic energy is \begin{equation} 
U_{B,rim}+U_{S,rim} \sim hYd^2\sqrt{h/R}. \label{total_energy_rim}
\end{equation}
Taking the derivative of the total elastic energy with respect to $ d $ and equating the result to the pushing force $ F $, we find $ d \sim (R/h)^{3/2}F/(YR) $. Thus $ C_{rr} \sim d/b^2 \sim (R/h)^{5/2}F/(YR^3) $, and $ |C_{rr}/C_{\theta\theta}| \sim (R/h)^{5/2}F/(YR^2)$. All the scaling relations obtained here agree with the numerical results presented in the previous section.

\subsubsection{Analytical solutions for c-cones}
The deformation of shells of revolution under symmetrical loading is a classical problem \cite {love1944,turner1965, gibson1965,flugge1960}. The governing F\"{o}ppl-von K\'{a}rm\'{a}n equations\cite{landau1} are 4th order nonlinear differential equations. The analysis is greatly simplified by treating the limiting regime of weak loading, so that the equations can be linearized in the deformation. From the linearized solution, we may then verify that for loading forces of interest, the linearized treatment is completely valid for asymptotically thin sheets.

For a c-cone, assuming that the supporting container's rim is infinitely hard (i.e. the range of the normal force from the rim is close to zero), $ \phi $ satisfies the following differential equations as shown in Appendix B:
\begin{equation}
r^2\frac{d^4\phi}{dr^4}+4r\frac{d^3\phi}{dr^3}+A_1\phi=
\begin{cases}
-\frac{A_1A_2}{r}  &\text{for $ r<R\csc(\theta_0) $} \\
0 &\text{for $ r> R \csc(\theta_0)$}
\end{cases}\quad ,
\label{phi_partial}
\end{equation}
where $ \theta_0 $ is half of the underlying cone's opening angle, $ A_1=12(1-\nu^2)\cot^2(\theta_0)/h^2 $, and $ A_2=F\sec^2(\theta_0)/(2\pi Yh) $.

These equations have closed form solutions in terms of Bessel functions. For $ r> R \csc(\theta_0) $, the equation is homogeneous and its general solution is \cite {love1944,turner1965}
\begin{equation}
\begin{split}
\phi(r)=&B_1[bei(\zeta)+\frac{2}{\zeta}ber'(\zeta)]+B_2[ber(\zeta)-\frac{2}{\zeta}bei'(\zeta)] \\
&+B_3[kei(\zeta)+\frac{2}{\zeta}ker'(\zeta)]+B_4[ker(\zeta)-\frac{2}{\zeta}kei'(\zeta)] 
\end{split}\quad,
\end{equation}
where $ \zeta=2 \sqrt[4]{12(1-\nu^2)r^2\tan^2(\theta_0)/h^2}  $, and a prime is differentiation with respect to $ \zeta $. The $ ber $, $ bei $, $ ker $, and $ kei $ functions are known as the Thomson or Kelvin functions \cite{watson1944,turner1965}. Evidently $ \phi $ goes to zero for large $ r $. However, $ bei $ and $ ber $ diverges there, and are linearly independent. Thus their coefficients $ B_1 $ and $ B_2 $ must vanish.
The solution for the $ r<R\csc(\theta_0) $ has an extra term:
\begin{equation}
\begin{split}
\phi(r)=&B_5[bei(\zeta)+\frac{2}{\zeta}ber'(\zeta)]+B_6[ber(\zeta)-\frac{2}{\zeta}bei'(\zeta)] \\
&+B_7[kei(\zeta)+\frac{2}{\zeta}ker'(\zeta)]+B_8[ker(\zeta)-\frac{2}{\zeta}kei'(\zeta)] \\
&-A_2/r
\end{split}\quad.
\end{equation}
Even though $ kei(\zeta) $ and $ ker(\zeta) $ are divergent at $ \zeta=0 $, the $ kei $ and $ ker $ terms are needed to balance the $ A_2/r $ term that also diverges at the apex. However, $ kei(\zeta) $ and $ ker(\zeta)$ decreases almost exponentially as $ \zeta $ increases, so these two terms are negligible near the rim. For our purpose of getting the radial profile of $ C_{rr} $ near the rim, the term $ -A_2/r $ can also be ignored because it changes in length scale $ \sim R $, which is much larger than the width of the rim region. The contribution of this term to $ C_{rr} $ at the rim  is vanishingly small compared with the $ C_{rr} $ we got from numerical simulations and scaling arguments:
\begin{equation}
\left[\frac{d}{dr}(-A_2/r)/C_{rr}\right]_{r=R\csc(\theta_0)} \sim \frac{A_2}{R^2}\frac{YR^3}{(R/h)^{5/2}F} \sim (\frac{h}{R})^{3/2}.
\end{equation}

We conclude from the preceding reasoning that it is sufficient to specify $ B_3 $, $ B_4 $, $ B_5 $, and $ B_6 $ for the inner and outer regions. To determine these four coefficients, we may use the four matching conditions applicable at the forcing point $ r=R\csc(\theta_0)$: $ \phi $ is continuous and equal to zero , the curvature $ {d\phi}/{dr} $ is also continuous, and there is a jump for $ {d^2\phi}/{d^2r} $ which equals $ -F/(2\pi\sin(\theta_0)\kappa R) $. The last condition is due to the assumption that the rim of the supporting container is infinitely hard. Any localized force on an elastic sheet produces a discontinuity of curvature derivative of this type \cite{landau1}.

\begin{figure}[h!]
  \begin{center}
  \includegraphics[width=0.5\textwidth]{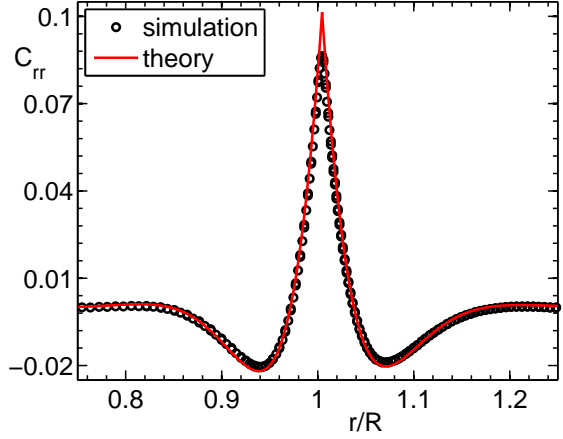}
  \caption{(Color online) Radial profiles of $C_{rr}$ in a c-cone from both numerical simulations and analytical solutions. $ C_{rr} $ is in units of $ 1/R $. The peak value from simulations is about $ 15\% $ lower than that from analytical solutions. The sheet here is the thinnest used in simulations with relative thickness $ h/R =0.00029 $. }
  \label{ccone_vs_theory}
  \end{center}
\end{figure}

Finally, we can compare the radial profile of $ C_{rr} $ from analytical solution with that from numerical simulation. The overall excellent agreement between them, as shown in Figure \ref{ccone_vs_theory}, gives convincing evidence that our numerical results are valid. It should be stressed that there is no normalization in either axis. The $ 15\% $ difference between the peak values are primarily due to three factors. First, as noted in Section II the percentage uncertainty of the rim curvatures from simulation is about $ 11\% $. Second, the analytical solution assumes an infinitely sharp container edge, but the force range of the normal force used in the simulation is close to $ 7\% $ of the full-width-at-half-maximum (FWHM) of the peak for this specific c-cone. Third, the local lattice spacing in the radial direction is also about $ 7\% $ of the FWHM. The last two factors cause the simulated curve to have a rounded peak. The effect of the finite thickness of the sheet is negligible here since the thickness is less than a tenth of the normal force range. Finite size of the simulation can also influence the curvature profile.

In addition to the stronger peak, the analytical solution shown in Figure \ref{ccone_vs_theory} has a $ 10\% $ stronger dip on either side. We believe this is due to the local compensation of Gaussian curvature \cite{wang1}, which requires the integral under the curve to be zero. So if the analytical solution has extra area under the peak, it must have extra negative area at the dips. 

Over the full range of the c-cone sheet thickness covered in our simulation, the peak values of the $ C_{rr} $ at the rim from simulation is lower than that given by the analytical solutions by between $ 10\% $ and $ 15\% $. We should expect a similar level of discrepancy between the simulation and analytical solutions for the d-cone. Thus the lowest $ |C_{rr}/C_{\theta\theta}|$ value achieved for d-cones in our simulations may increase from $ 0.76 $ to a value as high as $ 0.87 $, which is much closer to $ 1 $, but this level of discrepancy should have no material impact on the scaling relationship between $ |C_{rr}/C_{\theta\theta}|$, $ h $, and $ F $, which is the much stronger evidence that $ |C_{rr}/C_{\theta\theta}|$ at the rim will drop below one and keep decreasing as the thickness of the sheet approaches zero.

\section{DISCUSSION AND CONCLUSION}
In this paper we have shown numerically that contrary to previous claims, the mean curvature at the rim in a d-cone does not vanish as the thickness of the sheet goes to zero. This vanishing requires that $ |C_{rr}/C_{\theta\theta}| $ goes to $ 1 $. However, in the range we studied, $ |C_{rr}/C_{\theta\theta}| $ at the rim appears to vary as $ (h/R)^{1/3} $. More generally, in both c-cones and d-cones, $ |C_{rr}/C_{\theta\theta}| \sim (R/h)^{5/2}F/(YR^{2}) $.  These identical scaling laws and the similarity of the radial profiles of $ C_{rr} $ in d-cones and c-cones suggest that the core region of a d-cone has no influence on how the rim region reacts to the normal rim force pressure. The core region only affects the amplitude of the rim force pressure. 

This paper does not attempt to determine the right scaling law for the pushing force $ F $ in a d-cone. Simply combining the derived $ h^{-5/2}F $ scaling law with the numerically observed $ h^{1/3} $ scaling law for $ |C_{rr}/C_{\theta\theta}| $ will give $ F \sim Yh^{3}/R(h/R)^{-1/6} $. There is another proposed functional form for $ F $:  $ F \sim Yh^3/R \ln(R_p/R_c) $, where $ R_p $ is radius of the sheet, and $ R_c \sim h^{1/3}R^{2/3} $ is the radius of the core region \cite{cerda1,cerda2}. The first force scaling is asymptotically much stronger than the second one with the logarithmic term. However, within the dynamic range covered in our simulations, the fit for these two functional forms are equally good. It should also be mentioned that some researchers have expressed doubts on the arguments leading to the second functional form\cite{liang1, witten1}. Our work in progress aims to resolve this issue.

\textbf{Acknowledgments}

The author is grateful to Thomas A. Witten for enlightening discussions. He warmly thanks the University of Chicago Computer Science Instructional Laboratories for providing computing resources. This work was performed as part of the author's PhD research under the supervision of Thomas A. Witten and was supported in part by the US-Israel Binational Science Foundation and in part by the National Science Foundation's MRSEC Program under Grant Number DMR 0820054.

\appendix
\appendixpage
\addappheadtotoc
\section{Formulas for strain and curvature tensors}
\subsection{Formulas for the strain tensor}
This subsection derives the expression for strain tensor $ \gamma $ in terms of the changes of edge lengths $ \Delta d_1 $, $ \Delta d_2 $ and $ \Delta d_3 $ in an arbitrary triangle $ ABC $ as shown in Figure \ref{strain}(a) and \ref{strain}(b).

In Figure \ref{strain}(a), let the coordinates of $ A $, $ B $, and $ C $ be $ (0,0) $, $ (x_B,y_B) $, and $ (x_C,0) $, respectively. When triangle $ ABC $ is under strain, as shown in Figure \ref{strain}(b), denote the changes of $ x_B $, $ y_B $, and $ x_C $ as $ \Delta x_B $, $ \Delta y_B $, and $ \Delta x_C $, and the new coordinates of $ A $, $ B $, and $ C $ are $ (0,0) $, $ (x_B+\Delta x_B,y_B+\Delta y_B) $, and $ (x_C+ \Delta x_C,0) $. For a general point $ (x,y) $ in the triangle, let us denote its displacement as $ (u_x, u_y) $, so its coordinate due to deformation is $ (x+u_x, y+u_y) $. Under the assumption of infinitesimal constant strain across the triangle, $ \frac{\partial u_x}{\partial x}  $, $ \frac{\partial u_x}{\partial y} $, $ \frac{\partial u_y}{\partial x} $, and $ \frac{\partial u_y}{\partial y} $ should be constants. We can also expand the displacement of points $ B $ and $ C $ in terms of these partial derivatives: 
\begin{align}
\Delta x_B &=\frac{\partial u_x}{\partial x}x_B+\frac{\partial u_x}{\partial y}y_B + \text{higher order terms}\, ,\label{dxb}\\
\Delta y_B &=\frac{\partial u_y}{\partial x}x_B+\frac{\partial u_y}{\partial y}y_B + \text{higher order terms}\, ,\label{dyb}\\
\Delta x_C &=\frac{\partial u_x}{\partial x}x_C + \text{higher order terms}\, ,\label{dxc}\\
\Delta y_C &=\frac{\partial u_y}{\partial x}x_C+ \text{higher order terms}=0 \, .\label{dyc}
\end{align}
Ignoring the higher order terms, we can solve Equations \ref{dxb}, \ref{dyb},\ref{dxc},and \ref{dyc} for $ \frac{\partial u_x}{\partial x}  $, $ \frac{\partial u_x}{\partial y} $, $ \frac{\partial u_y}{\partial x} $, and $ \frac{\partial u_y}{\partial y} $:
\begin{align}
\frac{\partial u_x}{\partial x}&=\frac{\Delta x_C}{x_C} \, ,\\
\frac{\partial u_x}{\partial y}&=\frac{\Delta x_B}{y_B}-\frac{x_B\Delta x_C}{y_B x_C}\,,\\
\frac{\partial u_y}{\partial x}&=0\,, \\
\frac{\partial u_y}{\partial y}&=\frac{\Delta y_B}{y_B} \,.
\end{align}
Then we can easily get the strain tensor elements \cite{landau1}:
\begin{align}
\gamma_{xx}&=\frac{\partial u_x}{\partial x}=\frac{\Delta x_C}{x_C} \,, \\
\gamma_{yy}&=\frac{\partial u_y}{\partial y}=\frac{\Delta y_B}{y_B} \,, \\
\gamma_{xy}&=\gamma_{yx}=\frac{1}{2}(\frac{\partial u_x}{\partial y}+\frac{\partial u_y}{\partial x})=\frac{1}{2}(\frac{\Delta x_B}{y_B}-\frac{x_B\Delta x_C}{y_B x_C}) \,,
\end{align}
or in matrix form:
\begin{equation}
\left[
\begin{array}{c}
\gamma_{xx} \\ 
\gamma_{yy} \\ 
\gamma_{xy}
\end{array} 
\right]=S\left[
\begin{array}{c}
\Delta x_B \\ 
\Delta y_B \\ 
\Delta x_C
\end{array} 
\right] \label{gamma_delta_displacement} \,,
\end{equation}
where 
\begin{equation}
S=\left[
\begin{matrix}
0&0&\frac{1}{x_C} \\ 
0&\frac{1}{y_B}&0 \\ 
\frac{1}{2y_B}&0&-\frac{x_B}{2y_B x_C}
\end{matrix} 
\right] \label{s_expression} \,.
\end{equation}

We have determined the elements of $ \gamma $ in terms of $ \Delta x_B $, $ \Delta y_B $, and $ \Delta x_C $, but it is much more efficient to compute $ \Delta d_1 $, $ \Delta d_2 $, and $ \Delta d_3 $ during simulating, so let us find the expression for $ \Delta x_B $, $ \Delta y_B $, and $ \Delta x_C $ in terms of $ \Delta d_1 $, $ \Delta d_2 $, and $ \Delta d_3 $, and then express $ \gamma $ in terms of $ \Delta d_1 $, $ \Delta d_2 $, and $ \Delta d_3 $.

The change of length for edge $ AB $ is:
\begin{align}
\nonumber \Delta d_1 &=\sqrt{(x_B+\Delta x_B)^2+(y_B+\Delta y_B)^2}-\sqrt{x_B^2+y_B^2} \\ 
&=\frac{x_B}{\sqrt{x_B^2+y_B^2}}\Delta x_B+\frac{y_B}{\sqrt{x_B^2+y_B^2}}\Delta y_B+ \text{higher order terms} \, .
\label{delta_d1}
\end{align}
Similarly, we can express $ \Delta d_2 $ and $ \Delta d_3 $ in terms of $ \Delta x_B $, $ \Delta y_B $, and $ \Delta x_C $:
\begin{align}
\nonumber \Delta d_2 &=\frac{x_B-x_C}{\sqrt{(x_B-x_C)^2+y_B^2}}(\Delta x_B-\Delta x_C)+\frac{y_B}{\sqrt{(x_B-x_C)^2+y_B^2}}\Delta y_B \\
& \quad +\text{higher order terms} \, \label{delta_d2},\\
\Delta d_3 &=\Delta x_C \, . \label{delta_d3}
\end{align}
Let us ignore the higher order terms and rewrite Equations \ref{delta_d1}, \ref{delta_d2}, and \ref{delta_d3} in matrix form:
\begin{equation}
\left[
\begin{array}{c}
\Delta d_1 \\ 
\Delta d_2 \\ 
\Delta d_3
\end{array} 
\right]=G\left[
\begin{array}{c}
\Delta x_B \\ 
\Delta y_B \\ 
\Delta x_C
\end{array} 
\right] \label{delta_d_delta_x_relation} \, ,
\end{equation}
where
\begin{equation}
G=\left[
\begin{matrix}
\frac{x_B}{\sqrt{x_B^2+y_B^2}}&\frac{y_B}{\sqrt{x_B^2+y_B^2}}&0 \\ 
\frac{x_B-x_C}{\sqrt{(x_B-x_C)^2+y_B^2}}&\frac{x_C-x_B}{\sqrt{(x_B-x_C)^2+y_B^2}}&\frac{y_B}{\sqrt{(x_B-x_C)^2+y_B^2}} \\ 
0&0&1
\end{matrix} 
\right] \label{g_expression} \,.
\end{equation}

Combining Equations \ref{gamma_delta_displacement} and \ref{delta_d_delta_x_relation}:
\begin{equation}
\left[
\begin{array}{c}
\gamma_{xx} \\ 
\gamma_{yy} \\ 
\gamma_{xy}
\end{array} 
\right]=SG^{-1}\left[\begin{array}{c}
\Delta d_1 \\ 
\Delta d_2 \\ 
\Delta d_3
\end{array}\right] \,,
\end{equation}
where $ G^{-1} $ means the inverse of $ G $. Since both $ S $ and $ G $ depend only on $ x_B $, $ y_B $, and $ x_C $, the matrix $ SG^{-1} $ is determined by the initial geometry of the triangle and can be calculated at program initialization.

\subsection{Formulas for the curvature tensor}
To calculate the curvature tensor on an arbitrary triangle, e.g. triangle $ ABC $ in Figure \ref{bend_fig}, we can fit the coordinates of the six vertices of its three adjoining triangles to the following function \cite{didonna1, liang1}:
\begin{equation}
z_i=a_1+a_2 x_i+a_3 y_i+a_4 x_i^2+a_5 x_i y_i+a_6 y_i^2\,,\quad i=A\,,\dots \, ,F
\label{fit_paraboloid}
\end{equation}
where $ (x_i,y_i,z_i) $ are coordinates of the vertices in a local coordinate system. In this system, the $ z $ axis is perpendicular to $ ABC $ and its origin is at the center of $ ABC $. These choices ensure that $ a_2 $ and $ a_3 $ are negligible. Then the curvature tensor elements are determined through \cite{didonna1, liang1}
\begin{equation}
C_{xx}=2a_4\,,\quad C_{xy}=a_5\,,\quad C_{yy}=2a_6 \,.
\end{equation}
If we ignore the changes in the $ x_i $ and $ y_i $, the coefficient matrix in Equation \ref{fit_paraboloid} will stay the same during the simulation, which means for each triangle we only need to invert the coefficient matrix once at program initialization. Under the assumption that both bending and stretching are infinitesimal, this simplification will result in a second order error in the curvature tensor.

In this local coordinate system, $ z_A $, $ z_B $, and $ z_C $ are all zero by the choice of the $ z $ axis. The other three nonzero $ z $ coordinates $ z_D $, $ z_E $, and $ z_F $ can be determined from the three dihedral angles between $ ABC $ and its three adjoining neighbors. For example, $ z_D\approx d_{D,AC}\theta_1$, where $ d_{D,AC} $ is the distance from point $ D $ to edge $ AC $. So, alternatively, we can also determine the curvature tensor from the three dihedral angles $ \theta_1 $, $ \theta_2 $, and $ \theta_3 $.

This method works for both uniform and variable lattices.

\section{Analytical solutions for c-cones}

\begin{figure}[h!]
  \begin{center}
  \includegraphics[width=0.5\textwidth]{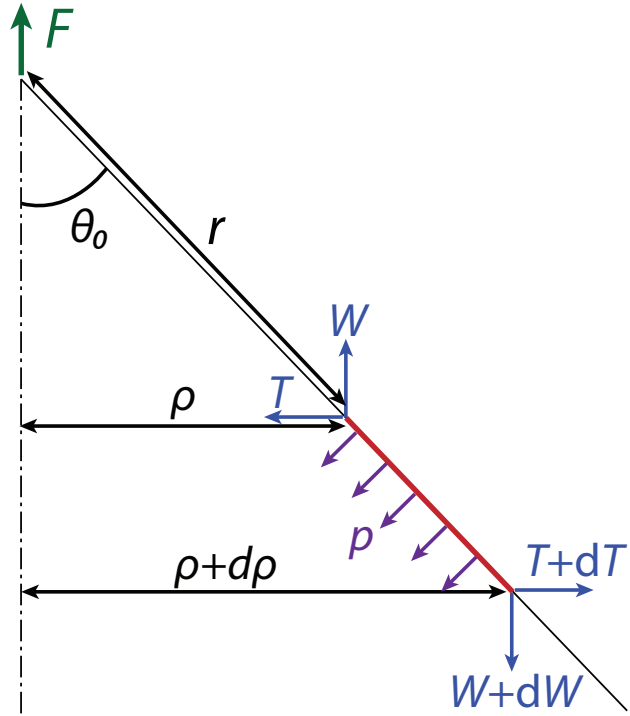}
  \caption{(Color online) Edge view of an element along the meridian in a c-cone. $ W $ and $ T $ are the axial and radial forces per unit length. $ \rho $ is the radial distance from the axis, and $ r $ is the meridional length (equivalent to the radial distance from the apex in the material coordinate). For an ideal c-cone, the normal pressure $ p $ is zero everywhere except at the rim. $ F $ and $ \theta_0 $ are the central pushing force and the half opening angle, respectively.}
  \label{element_mid_surface}
  \end{center}
\end{figure}

Ref. \cite{turner1965} provides detailed information on the general theory of symmetrically loaded shells of revolution, including conical shells. Its derivation assumes that the deformation is small relative to the size of the structure, but may be comparable to the thickness. In this regime, both bending and in-plane stretching may occur and thus need to be considered together. However, for many problems only the force and moment balance of the undistorted element is needed, and the resulting equations are normally linear. Here, we will simply quote the relevant equations presented in Ref. \cite{turner1965} and use them to find the specific equations for c-cones assuming that the supporting container's rim is infinitely hard.

As shown in Figure \ref{element_mid_surface}, $ p $, $ W $, and $ T $ are the pressure, axial and radial forces per unit length, respectively. $ \rho $ is the radial distance from the axis and it is related to $ r $ through $  \rho=r\sin(\theta_0) $. In a conical shell these variable and $ \phi $ satisfy the following equations according to Ref. \cite{turner1965}:
\begin{equation}
W\rho=-\int p\rho d\rho + V \, ,
\label{w_p_v}
\end{equation}
\begin{equation}
\kappa \sin(\theta_0)[\rho \frac{d}{d\rho}\{\frac{1}{\rho}\frac{d}{d\rho}(\rho \phi)\}]+T\rho\cot(\theta_0)=-\int p\rho d\rho +V\,,
\label{rho_T_general}
\end{equation}
\begin{equation}
\sin(\theta_0)[\rho \frac{d}{d\rho}\{\frac{1}{\rho}\frac{d}{d\rho}(\rho^2 T)\}]-hY\phi \cot(\theta_0) = \cos(\theta_0)[\frac{d}{d\rho}(p\rho^{2})-\frac{1}{\rho}\int p\rho d\rho + \frac{V}{\rho}]\,,
\label{T_rho_general}
\end{equation}
where $ V $ is a constant of integration. It is worth noting that these equations are derived for sheets whose unstressed (or undeformed) state has curvature, but we find that they are also valid when the unstressed state is flat.

Since we assume the supporting container's rim is infinitely hard, using the balance of force it is straightforward to determine that in c-cones $ W $ has the following functional form:
\begin{equation}
W=\frac{F}{2\pi \rho}H(R-\rho)\, 
\label{w}
\end{equation}
where $ H(x) $ is the Heaviside step function. Combining Equations \ref{w_p_v} and \ref{w}, we can get:
\begin{equation}
p=\frac{F}{2\pi R}\delta(\rho-R) \, , \text{and } V=\frac{F}{2\pi} \,,
\label{p_V}
\end{equation}
where $ \delta(x) $ is the Dirac delta function. Plugging this into Equations \ref{rho_T_general} and \ref{T_rho_general} and replacing $ \rho $ with $ r\sin(\theta_0) $ yield:
\begin{equation}
r \frac{d^2\phi}{dr^2}+\frac{d\phi}{dr}-\frac{\phi}{r}+\frac{Tr\cos(\theta_0)}{\kappa}= \begin{cases}
\frac{F}{2\pi \kappa } &\text{for } r <R\csc(\theta_0) \\
0 &\text{for } r >R\csc(\theta_0)
\end{cases} \, ,
\label{before_eliminating_T_1}
\end{equation}
\begin{equation}
r \frac{d^2(Tr)}{dr^2}+\frac{d(Tr)}{dr}-\frac{(Tr)}{r}-\frac{hY\phi\cos(\theta_0)}{\sin ^2(\theta_0)}= \begin{cases}
\frac{F\cos(\theta_0)}{2\pi r\sin^2(\theta_0)} &\text{for } r <R\csc(\theta_0) \\
0 &\text{for } r >R\csc(\theta_0)
\end{cases} \, .
\label{before_eliminating_T_2}
\end{equation}

Equation \ref{before_eliminating_T_1} can be rewritten to get an explicit expression for $ Tr $:
\begin{equation}
Tr=-\kappa \sec(\theta_0)[r \frac{d^2\phi}{dr^2}+\frac{d\phi}{dr}-\frac{\phi}{r}]+\begin{cases}
\frac{F\sec(\theta_0)}{2\pi } &\text{for } r <R\csc(\theta_0) \\
0 &\text{for } r >R\csc(\theta_0)
\end{cases} \, ,
\label{expression_Tr}
\end{equation}

Plugging Equation \ref{expression_Tr} into Equation \ref{before_eliminating_T_2}, we can get a forth order partial differential equation for $ \phi $:
\begin{equation}
r^2\frac{d^4\phi}{dr^4}+4r\frac{d^3\phi}{dr^3}+\frac{hY\phi}{\kappa}\cot^2(\theta_0)=\begin{cases}
-\frac{F\csc^2(\theta_0)}{2\pi r\kappa} &\text{for } r <R\csc(\theta_0) \\
0 &\text{for } r >R\csc(\theta_0)
\end{cases}
\end{equation}
which is equivalent to Equation \ref{phi_partial}.

\vspace*{-1ex} \bibliographystyle{apsrev}

\begin{thebibliography}{50}
\expandafter\ifx\csname natexlab\endcsname\relax\def\natexlab#1{#1}\fi
\expandafter\ifx\csname bibnamefont\endcsname\relax
  \def\bibnamefont#1{#1}\fi
\expandafter\ifx\csname bibfnamefont\endcsname\relax
  \def\bibfnamefont#1{#1}\fi
\expandafter\ifx\csname citenamefont\endcsname\relax
  \def\citenamefont#1{#1}\fi
\expandafter\ifx\csname url\endcsname\relax
  \def\url#1{\texttt{#1}}\fi
\expandafter\ifx\csname urlprefix\endcsname\relax\def\urlprefix{URL }\fi
\providecommand{\bibinfo}[2]{#2}
\providecommand{\eprint}[2][]{\url{#2}}

\bibitem[{\citenamefont{Cerda et~al.}(1999)\citenamefont{Cerda, Cha\"\i{}eb,
  Melo, and Mahadevan}}]{cerda2}
\bibinfo{author}{\bibfnamefont{E.}~\bibnamefont{Cerda}},
  \bibinfo{author}{\bibfnamefont{S.}~\bibnamefont{Cha\"\i{}eb}},
  \bibinfo{author}{\bibfnamefont{F.}~\bibnamefont{Melo}}, \bibnamefont{and}
  \bibinfo{author}{\bibfnamefont{L.}~\bibnamefont{Mahadevan}},
  \bibinfo{journal}{Nature} \textbf{\bibinfo{volume}{401}}, \bibinfo{pages}{46}
  (\bibinfo{year}{1999}).

\bibitem[{\citenamefont{Liang and Witten}(2006)}]{liang2}
\bibinfo{author}{\bibfnamefont{T.}~\bibnamefont{Liang}} \bibnamefont{and}
  \bibinfo{author}{\bibfnamefont{T.~A.} \bibnamefont{Witten}},
  \bibinfo{journal}{Phys. Rev. E} \textbf{\bibinfo{volume}{73}},
  \bibinfo{pages}{046604} (\bibinfo{year}{2006}).

\bibitem[{\citenamefont{Fasolino et~al.}(2007)\citenamefont{Fasolino, Los, and
  Katsnelson}}]{fasolino2007}
\bibinfo{author}{\bibfnamefont{A.}~\bibnamefont{Fasolino}},
  \bibinfo{author}{\bibfnamefont{J.~H.} \bibnamefont{Los}}, \bibnamefont{and}
  \bibinfo{author}{\bibfnamefont{M.~I.} \bibnamefont{Katsnelson}},
  \bibinfo{journal}{Nature Materials} \textbf{\bibinfo{volume}{6}},
  \bibinfo{pages}{858 } (\bibinfo{year}{2007}).

\bibitem[{\citenamefont{Cao et~al.}(2005)\citenamefont{Cao, Dickrell, Sawyer,
  Ghasemi-Nejhad, and Ajayan}}]{Cao25112005}
\bibinfo{author}{\bibfnamefont{A.}~\bibnamefont{Cao}},
  \bibinfo{author}{\bibfnamefont{P.~L.} \bibnamefont{Dickrell}},
  \bibinfo{author}{\bibfnamefont{W.~G.} \bibnamefont{Sawyer}},
  \bibinfo{author}{\bibfnamefont{M.~N.} \bibnamefont{Ghasemi-Nejhad}},
  \bibnamefont{and} \bibinfo{author}{\bibfnamefont{P.~M.}
  \bibnamefont{Ajayan}}, \bibinfo{journal}{Science}
  \textbf{\bibinfo{volume}{310}}, \bibinfo{pages}{1307} (\bibinfo{year}{2005}).

\bibitem[{\citenamefont{Arias and Arroyo}(2008)}]{arias2008}
\bibinfo{author}{\bibfnamefont{I.}~\bibnamefont{Arias}} \bibnamefont{and}
  \bibinfo{author}{\bibfnamefont{M.}~\bibnamefont{Arroyo}},
  \bibinfo{journal}{Phys. Rev. Lett.} \textbf{\bibinfo{volume}{100}},
  \bibinfo{pages}{085503} (\bibinfo{year}{2008}).

\bibitem[{\citenamefont{Lidmar et~al.}(2003)\citenamefont{Lidmar, Mirny, and
  Nelson}}]{lidmar1}
\bibinfo{author}{\bibfnamefont{J.}~\bibnamefont{Lidmar}},
  \bibinfo{author}{\bibfnamefont{L.}~\bibnamefont{Mirny}}, \bibnamefont{and}
  \bibinfo{author}{\bibfnamefont{D.~R.} \bibnamefont{Nelson}},
  \bibinfo{journal}{Phys. Rev. E} \textbf{\bibinfo{volume}{68}},
  \bibinfo{pages}{051910} (\bibinfo{year}{2003}).

\bibitem[{\citenamefont{Katifori et~al.}(2010)\citenamefont{Katifori, Alben,
  Cerda, Nelson, and Dumais}}]{Katifori27042010}
\bibinfo{author}{\bibfnamefont{E.}~\bibnamefont{Katifori}},
  \bibinfo{author}{\bibfnamefont{S.}~\bibnamefont{Alben}},
  \bibinfo{author}{\bibfnamefont{E.}~\bibnamefont{Cerda}},
  \bibinfo{author}{\bibfnamefont{D.~R.} \bibnamefont{Nelson}},
  \bibnamefont{and} \bibinfo{author}{\bibfnamefont{J.}~\bibnamefont{Dumais}},
  \bibinfo{journal}{PNAS} \textbf{\bibinfo{volume}{107}}, \bibinfo{pages}{7635}
  (\bibinfo{year}{2010}).

\bibitem[{\citenamefont{Pocivavsek et~al.}(2008)\citenamefont{Pocivavsek,
  Dellsy, Kern, Johnson, Lin, Lee, and Cerda}}]{Pocivavsek16052008}
\bibinfo{author}{\bibfnamefont{L.}~\bibnamefont{Pocivavsek}},
  \bibinfo{author}{\bibfnamefont{R.}~\bibnamefont{Dellsy}},
  \bibinfo{author}{\bibfnamefont{A.}~\bibnamefont{Kern}},
  \bibinfo{author}{\bibfnamefont{S.}~\bibnamefont{Johnson}},
  \bibinfo{author}{\bibfnamefont{B.}~\bibnamefont{Lin}},
  \bibinfo{author}{\bibfnamefont{K.~Y.~C.} \bibnamefont{Lee}},
  \bibnamefont{and} \bibinfo{author}{\bibfnamefont{E.}~\bibnamefont{Cerda}},
  \bibinfo{journal}{Science} \textbf{\bibinfo{volume}{320}},
  \bibinfo{pages}{912} (\bibinfo{year}{2008}).

\bibitem[{\citenamefont{Liang and Mahadevan}(2009)}]{Liang29122009}
\bibinfo{author}{\bibfnamefont{H.}~\bibnamefont{Liang}} \bibnamefont{and}
  \bibinfo{author}{\bibfnamefont{L.}~\bibnamefont{Mahadevan}},
  \bibinfo{journal}{PNAS} \textbf{\bibinfo{volume}{106}},
  \bibinfo{pages}{22049} (\bibinfo{year}{2009}).

\bibitem[{\citenamefont{Witten and Li}(1993)}]{witten2}
\bibinfo{author}{\bibfnamefont{T.~A.} \bibnamefont{Witten}} \bibnamefont{and}
  \bibinfo{author}{\bibfnamefont{H.}~\bibnamefont{Li}},
  \bibinfo{journal}{Europhys. Lett.} \textbf{\bibinfo{volume}{23}},
  \bibinfo{pages}{51} (\bibinfo{year}{1993}).

\bibitem[{\citenamefont{Lobkovsky et~al.}(1995)\citenamefont{Lobkovsky,
  Gentges, Li, Morse, and Witten}}]{lobkovsky3}
\bibinfo{author}{\bibfnamefont{A.}~\bibnamefont{Lobkovsky}},
  \bibinfo{author}{\bibfnamefont{S.}~\bibnamefont{Gentges}},
  \bibinfo{author}{\bibfnamefont{H.}~\bibnamefont{Li}},
  \bibinfo{author}{\bibfnamefont{D.}~\bibnamefont{Morse}}, \bibnamefont{and}
  \bibinfo{author}{\bibfnamefont{T.~A.} \bibnamefont{Witten}},
  \bibinfo{journal}{Science} \textbf{\bibinfo{volume}{270}},
  \bibinfo{pages}{1482} (\bibinfo{year}{1995}).

\bibitem[{\citenamefont{Lobkovsky and Witten}(1997)}]{lobkovsky2}
\bibinfo{author}{\bibfnamefont{A.~E.} \bibnamefont{Lobkovsky}}
  \bibnamefont{and} \bibinfo{author}{\bibfnamefont{T.~A.}
  \bibnamefont{Witten}}, \bibinfo{journal}{Phys. Rev. E}
  \textbf{\bibinfo{volume}{55}}, \bibinfo{pages}{1577} (\bibinfo{year}{1997}).

\bibitem[{\citenamefont{Ben~Amar and Pomeau}(1997)}]{amar1}
\bibinfo{author}{\bibfnamefont{M.}~\bibnamefont{Ben~Amar}} \bibnamefont{and}
  \bibinfo{author}{\bibfnamefont{Y.}~\bibnamefont{Pomeau}},
  \bibinfo{journal}{Roy. Soc. Lond. A} \textbf{\bibinfo{volume}{453}},
  \bibinfo{pages}{729} (\bibinfo{year}{1997}).

\bibitem[{\citenamefont{Cha\"\i{}eb et~al.}(1998)\citenamefont{Cha\"\i{}eb,
  Melo, and G\'eminard}}]{chaieb1}
\bibinfo{author}{\bibfnamefont{S.}~\bibnamefont{Cha\"\i{}eb}},
  \bibinfo{author}{\bibfnamefont{F.}~\bibnamefont{Melo}}, \bibnamefont{and}
  \bibinfo{author}{\bibfnamefont{J.-C.} \bibnamefont{G\'eminard}},
  \bibinfo{journal}{Phys. Rev. Lett.} \textbf{\bibinfo{volume}{80}},
  \bibinfo{pages}{2354} (\bibinfo{year}{1998}).

\bibitem[{\citenamefont{Cerda and Mahadevan}(1998)}]{cerda3}
\bibinfo{author}{\bibfnamefont{E.}~\bibnamefont{Cerda}} \bibnamefont{and}
  \bibinfo{author}{\bibfnamefont{L.}~\bibnamefont{Mahadevan}},
  \bibinfo{journal}{Phys. Rev. Lett.} \textbf{\bibinfo{volume}{80}},
  \bibinfo{pages}{2358} (\bibinfo{year}{1998}).

\bibitem[{\citenamefont{Cha\"\i{}eb and Melo}(1999)}]{chaieb2}
\bibinfo{author}{\bibfnamefont{S.}~\bibnamefont{Cha\"\i{}eb}} \bibnamefont{and}
  \bibinfo{author}{\bibfnamefont{F.}~\bibnamefont{Melo}},
  \bibinfo{journal}{Phys. Rev. E} \textbf{\bibinfo{volume}{60}},
  \bibinfo{pages}{6091} (\bibinfo{year}{1999}).

\bibitem[{\citenamefont{Boudaoud et~al.}(2000)\citenamefont{Boudaoud, Patricio,
  Couder, and Ben~Amar}}]{boudaoud1}
\bibinfo{author}{\bibfnamefont{A.}~\bibnamefont{Boudaoud}},
  \bibinfo{author}{\bibfnamefont{P.}~\bibnamefont{Patricio}},
  \bibinfo{author}{\bibfnamefont{Y.}~\bibnamefont{Couder}}, \bibnamefont{and}
  \bibinfo{author}{\bibfnamefont{M.}~\bibnamefont{Ben~Amar}},
  \bibinfo{journal}{Nature} \textbf{\bibinfo{volume}{407}},
  \bibinfo{pages}{718} (\bibinfo{year}{2000}).

\bibitem[{\citenamefont{Mora and Boudaoud}(2002)}]{mora1}
\bibinfo{author}{\bibfnamefont{T.}~\bibnamefont{Mora}} \bibnamefont{and}
  \bibinfo{author}{\bibfnamefont{A.}~\bibnamefont{Boudaoud}},
  \bibinfo{journal}{Europhys. Lett.} \textbf{\bibinfo{volume}{59}},
  \bibinfo{pages}{41} (\bibinfo{year}{2002}).

\bibitem[{\citenamefont{DiDonna}(2002)}]{didonna1}
\bibinfo{author}{\bibfnamefont{B.~A.} \bibnamefont{DiDonna}},
  \bibinfo{journal}{Phys. Rev. E} \textbf{\bibinfo{volume}{66}},
  \bibinfo{pages}{016601} (\bibinfo{year}{2002}).

\bibitem[{\citenamefont{Hamm et~al.}(2004)\citenamefont{Hamm, Roman, and
  Melo}}]{hamm1}
\bibinfo{author}{\bibfnamefont{E.}~\bibnamefont{Hamm}},
  \bibinfo{author}{\bibfnamefont{B.}~\bibnamefont{Roman}}, \bibnamefont{and}
  \bibinfo{author}{\bibfnamefont{F.}~\bibnamefont{Melo}},
  \bibinfo{journal}{Phys. Rev. E} \textbf{\bibinfo{volume}{70}},
  \bibinfo{pages}{026607} (\bibinfo{year}{2004}).

\bibitem[{\citenamefont{Cerda et~al.}(2004)\citenamefont{Cerda, Mahadevan, and
  Pasini}}]{Cerda17022004}
\bibinfo{author}{\bibfnamefont{E.}~\bibnamefont{Cerda}},
  \bibinfo{author}{\bibfnamefont{L.}~\bibnamefont{Mahadevan}},
  \bibnamefont{and} \bibinfo{author}{\bibfnamefont{J.~M.}
  \bibnamefont{Pasini}}, \bibinfo{journal}{PNAS}
  \textbf{\bibinfo{volume}{101}}, \bibinfo{pages}{1806} (\bibinfo{year}{2004}).

\bibitem[{\citenamefont{Farmer and Calladine}(2005)}]{Farmer2005509}
\bibinfo{author}{\bibfnamefont{S.}~\bibnamefont{Farmer}} \bibnamefont{and}
  \bibinfo{author}{\bibfnamefont{C.}~\bibnamefont{Calladine}},
  \bibinfo{journal}{International Journal of Mechanical Sciences}
  \textbf{\bibinfo{volume}{47}}, \bibinfo{pages}{509 } (\bibinfo{year}{2005}).

\bibitem[{\citenamefont{Blair and Kudrolli}(2005)}]{blair2005}
\bibinfo{author}{\bibfnamefont{D.~L.} \bibnamefont{Blair}} \bibnamefont{and}
  \bibinfo{author}{\bibfnamefont{A.}~\bibnamefont{Kudrolli}},
  \bibinfo{journal}{Phys. Rev. Lett.} \textbf{\bibinfo{volume}{94}},
  \bibinfo{pages}{166107} (\bibinfo{year}{2005}).

\bibitem[{\citenamefont{Liang and Witten}(2005)}]{liang1}
\bibinfo{author}{\bibfnamefont{T.}~\bibnamefont{Liang}} \bibnamefont{and}
  \bibinfo{author}{\bibfnamefont{T.~A.} \bibnamefont{Witten}},
  \bibinfo{journal}{Phys. Rev. E} \textbf{\bibinfo{volume}{71}},
  \bibinfo{pages}{016612} (\bibinfo{year}{2005}).

\bibitem[{\citenamefont{Cerda and Mahadevan}(2005)}]{cerda1}
\bibinfo{author}{\bibfnamefont{E.}~\bibnamefont{Cerda}} \bibnamefont{and}
  \bibinfo{author}{\bibfnamefont{L.}~\bibnamefont{Mahadevan}},
  \bibinfo{journal}{Proc. R. Soc. A} \textbf{\bibinfo{volume}{461}},
  \bibinfo{pages}{671} (\bibinfo{year}{2005}).

\bibitem[{\citenamefont{Andresen et~al.}(2007)\citenamefont{Andresen, Hansen,
  and Schmittbuhl}}]{andresen1}
\bibinfo{author}{\bibfnamefont{C.~A.} \bibnamefont{Andresen}},
  \bibinfo{author}{\bibfnamefont{A.}~\bibnamefont{Hansen}}, \bibnamefont{and}
  \bibinfo{author}{\bibfnamefont{J.}~\bibnamefont{Schmittbuhl}},
  \bibinfo{journal}{Phys. Rev. E} \textbf{\bibinfo{volume}{76}},
  \bibinfo{pages}{026108} (\bibinfo{year}{2007}).

\bibitem[{\citenamefont{Witten}(2007)}]{witten1}
\bibinfo{author}{\bibfnamefont{T.~A.} \bibnamefont{Witten}},
  \bibinfo{journal}{Rev. Mod. Phys.} \textbf{\bibinfo{volume}{79}},
  \bibinfo{pages}{643} (\bibinfo{year}{2007}).

\bibitem[{\citenamefont{Aharoni and Sharon}(2010)}]{aharoni2010}
\bibinfo{author}{\bibfnamefont{H.}~\bibnamefont{Aharoni}} \bibnamefont{and}
  \bibinfo{author}{\bibfnamefont{E.}~\bibnamefont{Sharon}},
  \bibinfo{journal}{Nature Materials} \textbf{\bibinfo{volume}{9}},
  \bibinfo{pages}{993} (\bibinfo{year}{2010}).

\bibitem[{\citenamefont{Mellado et~al.}(2011)\citenamefont{Mellado, Cheng, and
  Concha}}]{Mellado2011}
\bibinfo{author}{\bibfnamefont{P.}~\bibnamefont{Mellado}},
  \bibinfo{author}{\bibfnamefont{S.}~\bibnamefont{Cheng}}, \bibnamefont{and}
  \bibinfo{author}{\bibfnamefont{A.}~\bibnamefont{Concha}},
  \bibinfo{journal}{Phys. Rev. E} \textbf{\bibinfo{volume}{83}},
  \bibinfo{pages}{036607} (\bibinfo{year}{2011}).

\bibitem[{\citenamefont{Venkataramani}(2004)}]{Venkataramani2004}
\bibinfo{author}{\bibfnamefont{S.~C.} \bibnamefont{Venkataramani}},
  \bibinfo{journal}{Nonlinearity} \textbf{\bibinfo{volume}{17}},
  \bibinfo{pages}{301} (\bibinfo{year}{2004}).

\bibitem[{\citenamefont{Conti and Maggi}(2008)}]{conti2008}
\bibinfo{author}{\bibfnamefont{S.}~\bibnamefont{Conti}} \bibnamefont{and}
  \bibinfo{author}{\bibfnamefont{F.}~\bibnamefont{Maggi}},
  \bibinfo{journal}{Archive for Rational Mechanics and Analysis}
  \textbf{\bibinfo{volume}{187}}, \bibinfo{pages}{1} (\bibinfo{year}{2008}).

\bibitem[{\citenamefont{Landau and Lifshitz}(1986)}]{landau1}
\bibinfo{author}{\bibfnamefont{L.~D.} \bibnamefont{Landau}} \bibnamefont{and}
  \bibinfo{author}{\bibfnamefont{E.~M.} \bibnamefont{Lifshitz}},
  \emph{\bibinfo{title}{Theory of Elasticity}} (\bibinfo{publisher}{Pergamon
  Press}, \bibinfo{address}{New York}, \bibinfo{year}{1986}).

\bibitem[{\citenamefont{Lobkovsky}(1996)}]{lobkovsky1}
\bibinfo{author}{\bibfnamefont{A.~E.} \bibnamefont{Lobkovsky}},
  \bibinfo{journal}{Phys. Rev. E} \textbf{\bibinfo{volume}{53}},
  \bibinfo{pages}{3750} (\bibinfo{year}{1996}).

\bibitem[{\citenamefont{Mansfield}(1964)}]{mansfield1}
\bibinfo{author}{\bibfnamefont{E.~H.} \bibnamefont{Mansfield}},
  \emph{\bibinfo{title}{The Bending and Stretching of Plates}}
  (\bibinfo{publisher}{Pergamon}, \bibinfo{address}{New York},
  \bibinfo{year}{1964}).

\bibitem[{\citenamefont{Fraser}(1991)}]{fraser1991}
\bibinfo{author}{\bibfnamefont{C.~G.} \bibnamefont{Fraser}},
  \bibinfo{journal}{Centaurus} \textbf{\bibinfo{volume}{34}},
  \bibinfo{pages}{211} (\bibinfo{year}{1991}).

\bibitem[{\citenamefont{Wang and Witten}(2009)}]{wang1}
\bibinfo{author}{\bibfnamefont{J.~W.} \bibnamefont{Wang}} \bibnamefont{and}
  \bibinfo{author}{\bibfnamefont{T.~A.} \bibnamefont{Witten}},
  \bibinfo{journal}{Phys. Rev. E} \textbf{\bibinfo{volume}{80}},
  \bibinfo{pages}{046610} (\bibinfo{year}{2009}).

\bibitem[{\citenamefont{Pogorelov}(1988)}]{pogorelov1988}
\bibinfo{author}{\bibfnamefont{A.}~\bibnamefont{Pogorelov}},
  \emph{\bibinfo{title}{Bendings of surfaces and stability of shells}},
  Translations of mathematical monographs (\bibinfo{publisher}{American
  Mathematical Societ}, \bibinfo{address}{Providence, R.I},
  \bibinfo{year}{1988}).

\bibitem[{\citenamefont{Audoly}(1999)}]{audoly1999}
\bibinfo{author}{\bibfnamefont{B.}~\bibnamefont{Audoly}},
  \bibinfo{journal}{Phys. Rev. Lett.} \textbf{\bibinfo{volume}{83}},
  \bibinfo{pages}{4124} (\bibinfo{year}{1999}).

\bibitem[{\citenamefont{Jin and Kohn}(2000)}]{jin2000}
\bibinfo{author}{\bibfnamefont{W.}~\bibnamefont{Jin}} \bibnamefont{and}
  \bibinfo{author}{\bibfnamefont{R.}~\bibnamefont{Kohn}},
  \bibinfo{journal}{Journal of Nonlinear Science}
  \textbf{\bibinfo{volume}{10}}, \bibinfo{pages}{355} (\bibinfo{year}{2000}).

\bibitem[{\citenamefont{Struik}(1961)}]{struik1}
\bibinfo{author}{\bibfnamefont{D.~J.} \bibnamefont{Struik}},
  \emph{\bibinfo{title}{Lectures on Classical Differential Geometry}}
  (\bibinfo{publisher}{Addison-Wesley Publishing Company},
  \bibinfo{address}{Massachusetts}, \bibinfo{year}{1961}).

\bibitem[{\citenamefont{Love}(1944)}]{love1944}
\bibinfo{author}{\bibfnamefont{A.}~\bibnamefont{Love}}, \emph{\bibinfo{title}{A
  Treatise on the Mathematical Theory of Elasticity}}
  (\bibinfo{publisher}{Dover Publications}, \bibinfo{address}{New York},
  \bibinfo{year}{1944}), \bibinfo{edition}{4th} ed.

\bibitem[{\citenamefont{Turner}(1965)}]{turner1965}
\bibinfo{author}{\bibfnamefont{C.~E.} \bibnamefont{Turner}},
  \emph{\bibinfo{title}{Introduction to Plate and Shell Theory}}
  (\bibinfo{publisher}{American Elsevier Publishing Company},
  \bibinfo{year}{1965}).

\bibitem[{aba(2009)}]{abaqus2009}
\emph{\bibinfo{title}{Abaqus Theory Manual (6.9)}},
  \bibinfo{organization}{Dassault Systèmes Simulia Corp.},
  \bibinfo{address}{Providence, RI} (\bibinfo{year}{2009}).

\bibitem[{\citenamefont{Wang and Witten}()}]{wang2}
\bibinfo{author}{\bibfnamefont{J.~W.} \bibnamefont{Wang}} \bibnamefont{and}
  \bibinfo{author}{\bibfnamefont{T.~A.} \bibnamefont{Witten}},
  \bibinfo{note}{manuscript in preparation}.

\bibitem[{\citenamefont{Seung and Nelson}(1988)}]{seung1}
\bibinfo{author}{\bibfnamefont{H.~S.} \bibnamefont{Seung}} \bibnamefont{and}
  \bibinfo{author}{\bibfnamefont{D.~R.} \bibnamefont{Nelson}},
  \bibinfo{journal}{Phys. Rev. A} \textbf{\bibinfo{volume}{38}},
  \bibinfo{pages}{1005} (\bibinfo{year}{1988}).

\bibitem[{\citenamefont{Press et~al.}(1996)\citenamefont{Press, Teukolsky,
  Vetterling, and Flannery}}]{press1}
\bibinfo{author}{\bibfnamefont{W.~H.} \bibnamefont{Press}},
  \bibinfo{author}{\bibfnamefont{S.~A.} \bibnamefont{Teukolsky}},
  \bibinfo{author}{\bibfnamefont{W.~T.} \bibnamefont{Vetterling}},
  \bibnamefont{and} \bibinfo{author}{\bibfnamefont{B.~P.}
  \bibnamefont{Flannery}}, \emph{\bibinfo{title}{Numerical Recipe in C}}
  (\bibinfo{publisher}{Cambridage University Press},
  \bibinfo{address}{Cambridage}, \bibinfo{year}{1996}).

\bibitem[{\citenamefont{J\'{a}nosi et~al.}(1998)\citenamefont{J\'{a}nosi,
  Chr\'{e}tien, and Flyvbjerg}}]{janosi1998}
\bibinfo{author}{\bibfnamefont{I.~M.} \bibnamefont{J\'{a}nosi}},
  \bibinfo{author}{\bibfnamefont{D.}~\bibnamefont{Chr\'{e}tien}},
  \bibnamefont{and}
  \bibinfo{author}{\bibfnamefont{H.}~\bibnamefont{Flyvbjerg}},
  \bibinfo{journal}{European Biophysics Journal} \textbf{\bibinfo{volume}{27}},
  \bibinfo{pages}{501} (\bibinfo{year}{1998}).

\bibitem[{\citenamefont{Gibson}(1965)}]{gibson1965}
\bibinfo{author}{\bibfnamefont{J.~E.} \bibnamefont{Gibson}},
  \emph{\bibinfo{title}{Linear Elastic Theory of Thin Shells}}
  (\bibinfo{publisher}{Pergamon Press}, \bibinfo{year}{1965}).

\bibitem[{\citenamefont{Fl\"{u}gge}(1960)}]{flugge1960}
\bibinfo{author}{\bibfnamefont{W.}~\bibnamefont{Fl\"{u}gge}},
  \emph{\bibinfo{title}{Stresses in Shells}}
  (\bibinfo{publisher}{Springer-Verlag}, \bibinfo{address}{Berlin},
  \bibinfo{year}{1960}).

\bibitem[{\citenamefont{Watson}(1944)}]{watson1944}
\bibinfo{author}{\bibfnamefont{G.~N.} \bibnamefont{Watson}},
  \emph{\bibinfo{title}{A Treatise on the Theory of Bessel Functions}}
  (\bibinfo{publisher}{Cambridge University Press}, \bibinfo{year}{1944}),
  \bibinfo{edition}{2nd} ed.

\end{thebibliography}

\end{document}